\documentclass[fleqn,usenatbib]{mnras}

\usepackage{newtxtext,newtxmath}
\usepackage[T1]{fontenc}

\usepackage{graphicx}
\usepackage{booktabs}
\usepackage{multirow}
\usepackage[export]{adjustbox}
\usepackage{siunitx}

\usepackage[caption=false]{subfig}

\usepackage{xspace} % for \xspace used below
\newcommand{\hrieuv}{HRI\textsubscript{EUV}\xspace}

\title[Effects from Coronal Rain Impact]{Compression, Impact and Hot Rebound Flows from Coronal Rain Downflows}

\author[J. Wachira \& P. Antolin]{
J. Wachira,$^{1}$
P. Antolin,$^{1}$\thanks{E-mail: patrick.antolin@northumbria.ac.uk}
\\
$^{1}$School of Engineering, Physics and Mathematics, Northumbria University, Newcastle upon Tyne, NE1 8ST, UK
}

\date{Accepted XXX. Received YYY; in original form ZZZ}
\pubyear{\the\year}
\begin{document}
\label{firstpage}
\pagerange{\pageref{firstpage}--\pageref{lastpage}}
\maketitle

\begin{abstract} %250 word limit
Studying coronal rain formation through thermal non-equilibrium (TNE) and thermal instability (TI) provides insights into coronal heating mechanisms. We {analysed a quiescent coronal rain event using }space-based {observations} from the High-Resolution Imager in Extreme Ultraviolet (\hrieuv) of Solar Orbiter (SolO), the {Atmospheric Imaging Assembly} (AIA) of the Solar Dynamics Observatory (SDO), and the Slit-Jaw Imager (SJI) from the {Interface Region Imaging Spectrograph} (IRIS) from November 1st, 2023. During the coronal rain shower, the coronal loop exhibits {substantial} EUV variability and {structural} changes. Rain clumps {fell at} $72-87$~km~s$^{-1}$ with cool EUV absorbing core sizes of $\approx$600~km and densities of $\approx6\times10^{11}$~cm$^{-3}$ {preceded by} strong compressions. {These mostly isothermal compressions suggest energy transfer into the rain, decelerating it and possibly reducing cooling rates -- consistent with accretion braking timescales}. The shower carried {microflare-level energy ($4.64\times10^{26}$~erg)}, with clumps producing impacts that reach the lower transition region and are {visible across} all EUV channels and in SJI~1400~\AA. The impacts generated hot rebound flows ($10^{6.2}-10^{6.3}~$K, $85-87$ km~s$^{-1}$) that refilled and reheated the loop but carried less than {$15\%$} of the clumps' kinetic energy. We {detected} steady footpoint heating {signatures consistent} with the TNE-TI scenario, with an estimated amplitude of {$10^{-2\pm0.3}~$} erg~cm$^{-3}$ ~s$^{-1}$ {and heating scale heights of $2-10$~Mm, matching} active region {values}. Coronal rain may {thus serve as both a template for accretion braking and }a proxy for integrated heating {driving} TNE-TI {cycles}. 
\end{abstract}

%TC:ignore


\section{Introduction} \label{sec:introduction}

%Until the 1940s, it was widely accepted that temperatures within the sun decrease radially outwards. However, this viewpoint was called into question and a discovery \citep{grotrian1934fraunhofersche,edlen1945identification} was made indicating that the solar corona reheats to temperatures exceeding millions of degrees Kelvin. 
The heating mechanism of coronal loops, {the building blocks of the inner solar corona}, has eluded scientists for many years to what is now known as the `coronal heating problem'. 
%Through the help of countless observations and numerical simulations \citep{van2020coronal,van2014alfven,dere1994explosive} it is evident that the heating mechanism involves the release of magnetic energy through processes such as magnetic reconnection and magnetohydrodynamic (MHD) waves. 
Coronal loops consist of loop-like magnetic field lines that are anchored to the solar surface and contain ionised plasma. The plasma is frozen-in to the magnetic field lines due to the high magnetic Reynolds number. Observations \citep{reale2014coronal} have identified filamentary substructures within coronal loops, leading to the concept known as coronal strands \citep{klimchuk2015key}. The widths of these coronal strands have been measured to be a few hundred kilometres \citep{williams2020high, peter2013structure,brooks2013high,aschwanden2011solar}. Not all coronal loops show coronal strands, and it has been proposed that strands appear during the coronal loops cooling phase, and particularly the formation of coronal rain, due to the condensation-corona transition region (CCTR), as observed by \citet{antolin2023extreme} in high-resolution with EUI$_{\hrieuv}$ of Solar Orbiter (SolO) \citep{ Muller2020, Rochus2020}, which is also supported by numerical simulations \citep{Antolin_2022ApJ...926L..29A}.

Coronal rain is characterised as cool, dense and clumpy plasma that exhibits temperature ranges from $10^3-10^{5.5}$\,~K and number density ranges of $10^{10}-10^{13}$\, cm$^{-3}$. Coronal rain appears over the order of minutes and is most easily seen when observed off the limb, when it is seen in emission in chromospheric lines such as H$\alpha$ and \ion{Ca}{II}~H or in transition region lines such as \ion{Si}{IV}~1402\,\AA. Studies by \citet{antolin2012observing, Ahn_2014SoPh..289.4117A, csahin2023spatial} have observed coronal rain to have a broad velocity distribution of $10 - 150$\, km\, s$^{-1}$, {with terminal velocities} peaking between $50-100$\, km\, s$^{-1}$, {and with accelerations about a third of solar gravity, and half that expected from effective gravity along semi-circular loops \citep{Schrijver_2001SoPh..198..325S,DeGroof05}}. The widths vary depending on the spatial resolution and the opacity of the spectral line of observation, averaging around {$50-500$}~km in the optical spectrum and $500-600$~km in the UV and EUV \citep{Tamburri_2025ApJ...990L...3T,antolin2015multi, antolin2023extreme, csahin2023spatial}. Rain lengths also exhibit a dependence but at a minor degree on the wavelength of observation, and are more sparsely distributed than the widths. Lengths in the optical spectra peak around $700-1000$~km, but about twice this value in the UV and low-temperature EUV lines. However, a long distribution tail is observed to higher length values for all wavelengths. For a review of the coronal rain properties please see \citet{antolin2022multi}.

The recent study by \citet{antolin2023extreme} identified an atmospheric response to the rain impact. They found speeds of the upward propagating features following the rain impact in the range of $50 - 130$\, km\, s$^{-1}$, and interpreted the lower and upper values in this range as corresponding to upflows and rebound shocks, respectively. This atmospheric response associated with coronal rain has also been predicted in numerical studies, e.g. by \citet{muller2003dynamics}.

Neighboring coronal rain clumps are commonly observed to fall in tandem with each other, spanning over a wide cross-section of coronal loops. This phenomenon is known as a `rain shower' \citep{antolin2012observing, Sahin_2022ApJ...931L..27S}. Local thermal instability (TI) within a coronal loop in a state of thermal-non-equilibrium (TNE) is said to be the cause of coronal rain, and is also thought to be the synchronising mechanism leading to rain showers. This is supported by numerical simulations by \citet{Fang_2013ApJ...771L..29F}.  TNE occurs when there is strongly stratified and relatively long-lasting heating. Namely, the loop footpoints need to be heated roughly ten times more than the loop apex, without much heating asymmetry between the two footpoints and the heating needs to be sustained longer than the radiative cooling time \citep{klimchuk2019role, Johnston_2019AA...625A.149J}. After a rapid heating phase, the coronal loop becomes increasingly dense due to sustained chromospheric evaporation, which in turn leads to an over-density compared to that needed to achieve thermal equilibrium. Radiative losses then become larger than the input from thermal conduction and enthalpy, leading to a runaway cooling and localised TI in the corona \citep{Antolin_2020PPCF...62a4016A,Keppens_2025ApJ...989...51K}. Condensations occur, which fall as coronal rain largely evacuating the loops. This TNE-TI scenario repeats if the heating persists, leading to TNE-TI cycles of heating and cooling that manifest as long-period EUV intensity pulsations \citep{Auchere_2014AA...563A...8A, Froment_2015ApJ...807..158F} and periodic coronal rain \citep{Auchere_2018ApJ...853..176A}. %However, there is still an ongoing debate on the role of TI, and the exact route to catastrophic cooling \citep{Klimchuk_2019, Waters_2024arXiv240815869W}. 

\citet{Antolin_2022ApJ...926L..29A} conducted 2.5D radiative MHD simulations which depict TI-driven catastrophic cooling {that} leads to the formation of cool and dense cores at the head of the rain emitting in chromospheric lines. These cores are surrounded by thin but strongly emitting shells in TR lines (the CCTR) that elongate in the wake of the rain.

{This work focuses on the coronal rain observed in non-flaring active regions (ARs), known as quiescent coronal rain. }
%There are three different types of coronal rain \citep{antolin2022multi}. The first is the quiescent type usually observed in non-flaring ARs and does not rely on a specific magnetic topology. This study focuses on this type. The second is flare-driven coronal rain which is associated with the cool material observed in the gradual phase of flare loops. Lastly, there is a prominence-coronal rain hybrid, also known as coronal spider or cloud prominence, which involves a complex magnetic field with a fan-spine (null-point) topology at the top of loop arcades. 
With the help of the {Atmospheric Imaging Assembly} \citep[AIA,][]{Lemen_2012SoPh..275...17L} onboard the Solar Dynamics Observatory \citep[SDO,][]{Pesnell_2012SoPh..275....3P}, alongside the Slit-Jaw Imager (SJI) onboard the {Interface Region Imaging Spectrograph} \citep[IRIS,][]{DePontieu_2014SoPh..289.2733D}, and the {High-Resolution Imager of the Extreme Ultra Violet} Instrument (\hrieuv) aboard the Solar Orbiter (SolO) mission \citep{Rochus2020, Muller2020}, this study provides a detailed analysis of quiescent coronal rain observed during the SolO perihelion of November 1st 2023. 
Through the variety of wavelengths and viewpoints, we examined various properties of coronal rain. In particular, we focus on the compression produced by the rain and its effect on the coronal loop upon impact on the lower atmosphere. In section \ref{sec:observations} we discuss our observation, in section~\ref{sec:methodology} we discuss our methodology, in section~\ref{sec:results} we discuss our findings, and a detailed analysis is discussed in section~\ref{sec:Discussion and Conclusion}.

\section{Observations}\label{sec:observations}

We sourced data from the 174~\AA\, passband of the High-Resolution Imager in Extreme Ultraviolet (\hrieuv) instrument aboard Solar Orbiter \citep[SolO,][]{Rochus2020}, as well as from the {Atmospheric Imaging Assembly} \citep[AIA,][]{Lemen_2012SoPh..275...17L} aboard the Solar Dynamics Observatory (SDO) and the Slit-Jaw Imager (SJI) onboard the {Interface Region Imaging Spectrograph} \citep[IRIS,][]{DePontieu_2014SoPh..289.2733D}. To identify the same region between AIA and the \hrieuv data, we utilised the JHelioviewer software developed by \citet{muller2017jhelioviewer}.

\begin{figure}
    \centering
    \includegraphics[width=0.7\columnwidth]{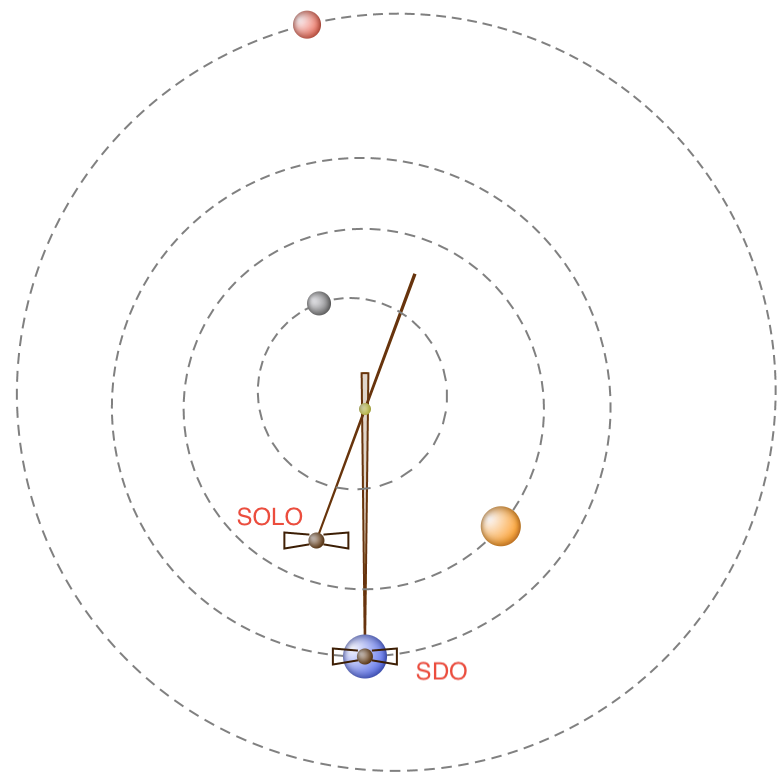}
    \caption{The location of SolO and SDO in relation to the Sun and Earth on November 1st, 2023, obtained from the Propagation Tool software \citep{rouillard2017propagation}. }
    \label{fig:Propagation Tool}
\end{figure}

\hrieuv was positioned at 0.56 AU from the Sun and captured images in a format of $2048 \times 2048$ pixels. This provided a full field-of-view (FOV) of $1007\arcsec.6 \times 1007\arcsec.6$, with a pixel scale of 0\arcsec.492 per pixel and a pixel sampling of 199.83\,km\, per pixel. The observation began on the 1st of November 2023 at 00:43:41 UT, with the last image recorded at 01:43:31 UT. An extra 3.65 minutes was added to the start and end times to account for the time it takes light to reach the AIA compared to \hrieuv. This observation was part of the `R SMALL HRES MCAD AR-Heating’ Solar Orbiter Planning Sheet (SOOP) which provided a cadence of 10~s, an exposure time of 3~s and a total duration of approximately 60~min. In the 174~\AA\, passband, \ion{Fe}{IX} (171~\AA) and \ion{Fe}{X} (174~\AA)  are the ions that dominate with a peak formation temperature of $10^6~$K. The images have been prepped with the \texttt{euiprep} routine to level 2, which reduces jitter and pointing errors. Significant jitter remained and a cross-correlation technique was applied to remove this, following  \citet{chitta2022solar}. {We have estimated the accuracy of this algorithm by randomly adding or subtracting a 1-$\sigma$ errors (photon shot noise) to each pixel and running the cross-correlation algorithm again. Repeating this process 100 times has yielded a standard deviation on the displacement shifts on the order of 10~km.}

IRIS and AIA level 2 data were obtained from the LMSAL data search website \href{https://iris.lmsal.com/search/}{[online]}, with both instruments positioned at approximately 1 AU from the Sun. We analysed across multiple AIA passbands: 94~\AA, 131~\AA, 171~\AA, 193~\AA, 211~\AA, 304~\AA, 335~\AA, 1600~\AA\, ~and 1700~\AA. This broad spectrum allowed for a comprehensive analysis of the impact and effect of the rain on the Sun’s atmosphere. 

\begin{figure}
    \centering
    \includegraphics[width=1.0\columnwidth]{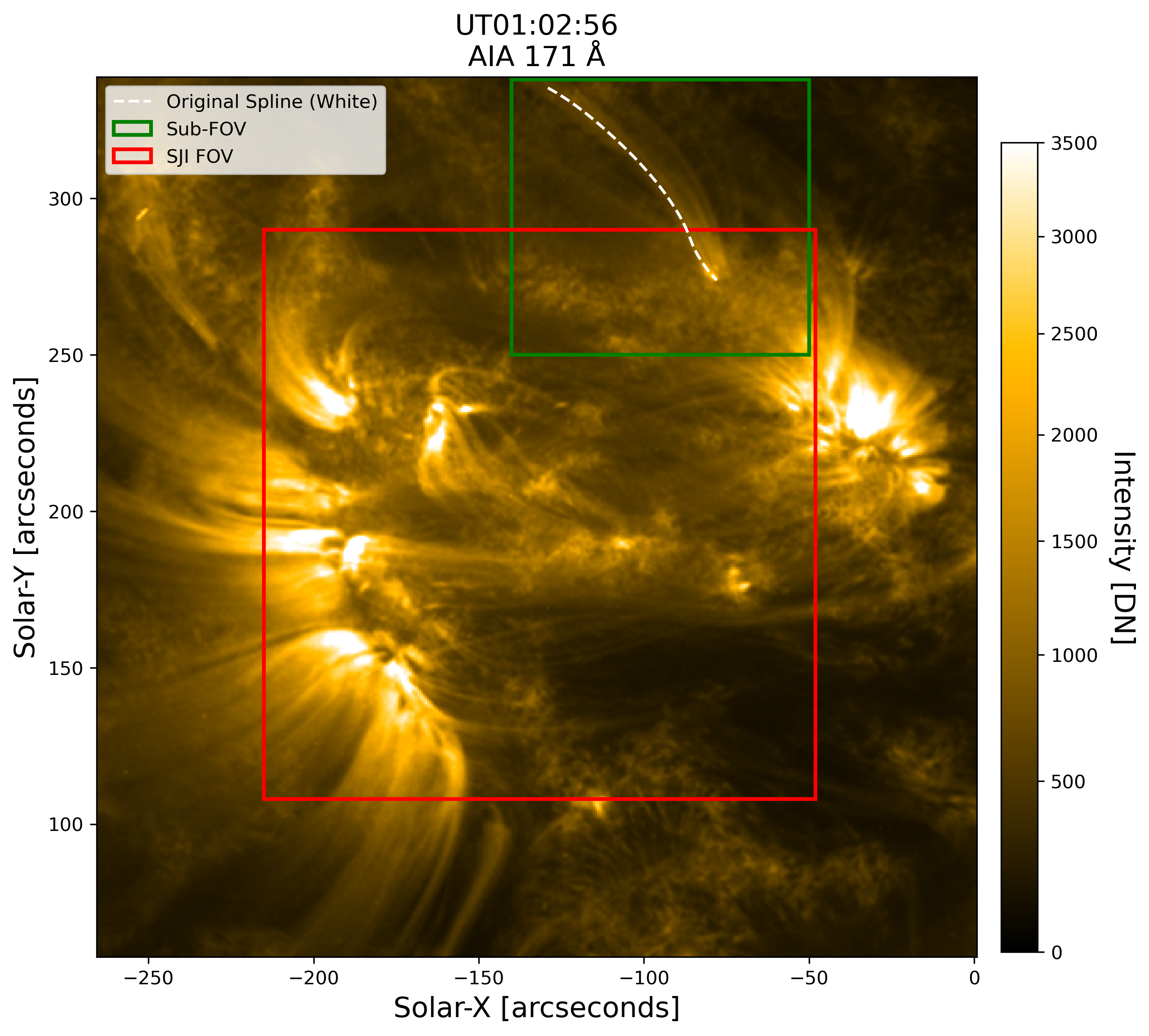}
    \caption{The active region observed in AIA showing the SJI 1400~\AA\, FOV and the sub-FOV shown in Figure \ref{fig:171_sub_fov}. The rain trajectory is plotted in white as `original spline'}
    \label{fig:171_full_fov}
\end{figure}

While most AIA passbands operated at a cadence of 12~s, AIA~1600~\AA\, ~and AIA~1700~\AA\, operated at a cadence of 24 seconds. AIA has a pixel scale of 0\arcsec.6 per pixel. The AIA dataset obtained for this study starts on the 1st of November at 00:29:26 UT and ends at 01:59:40 UT allowing for a total duration of roughly 90~min. The pixel sampling obtained from AIA is 431~km\, per pixel.

\begin{figure}
    \centering
    \includegraphics[width=1.0\columnwidth]{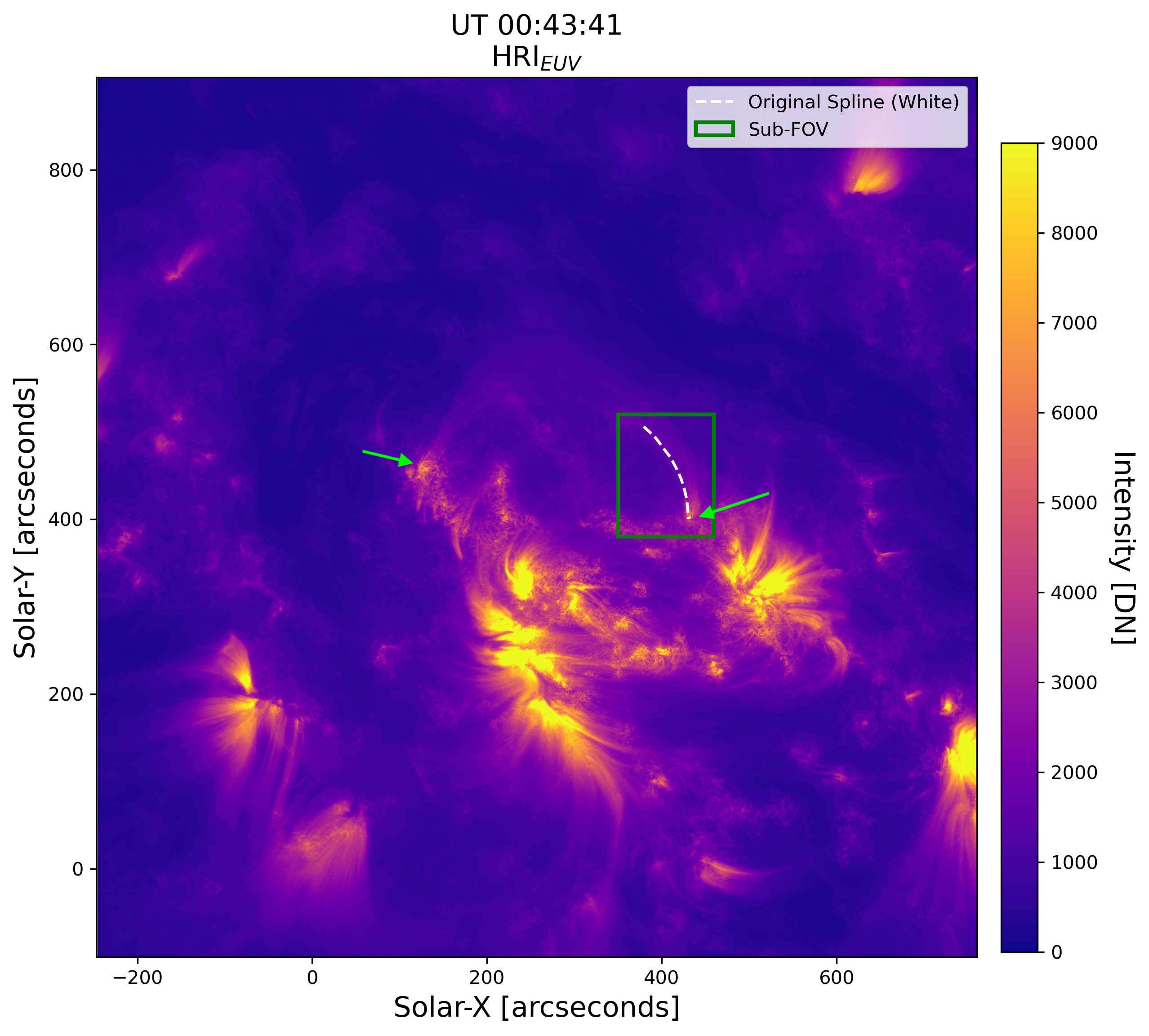}
    \caption{The studied AR on the solar disc with \hrieuv. The sub-FOV enclosed in green is shown in Figure \ref{fig:hri_sub_fov}. The rain trajectory is plotted in white and is denoted as the `original spline'. }
    \label{fig:hri_full_fov}
\end{figure}

We use SJI~1400~\AA\ {(OBS-ID 3440007611)}, data from IRIS, which predominantly captures the \ion{Si}{IV} resonance lines at 1394~\AA\, ~and 1402~\AA, which originate from plasma characteristic of the transition region and have a mean formation temperature of 0.08~MK. The SJI 1400\, \AA\, operated with a cadence of 10.5~s and an exposure time of 4~s. Each image obtained by SJI 1400~\AA\, has dimensions of $1001 \times 1093$ pixels with a pixel scale of 0\arcsec.166 per pixel. This provided a full field of view of $166\arcsec.52 \times 181\arcsec.82$ and a pixel sampling size of 120\, km\, per pixel. The FOV centre is -131.5 West and 199.24 North. The observation began on the 1st of November at 00:38:17 UTC and concluded at 01:49:49 UTC allowing for a total duration of approximately 70~min. IRIS conducted a very large sparse 2-step raster observation for this event. The slit did not cross the event and IRIS only party captured the loop footpoint as shown in Figure~\ref{fig:171_full_fov}.

The NOAA 13473 AR seen in Figure~\ref{fig:hri_full_fov} exhibited signs of being in its decay phase due to the lack of sunspots and pores as observed from the HMI continuum. Utilising the HMI magnetogram we note that the coronal loop footpoint, where the coronal rain is observed to fall, is anchored into a positive polarity region. The other footpoint anchored into a negative polarity region can also be distinguished. Both footpoints appear to be rooted in moss as depicted by the green arrows in \ref{fig:hri_full_fov}.

\section{Methodology}\label{sec:methodology}

\begin{figure}
    \centering
    \includegraphics[width=1.0\columnwidth]{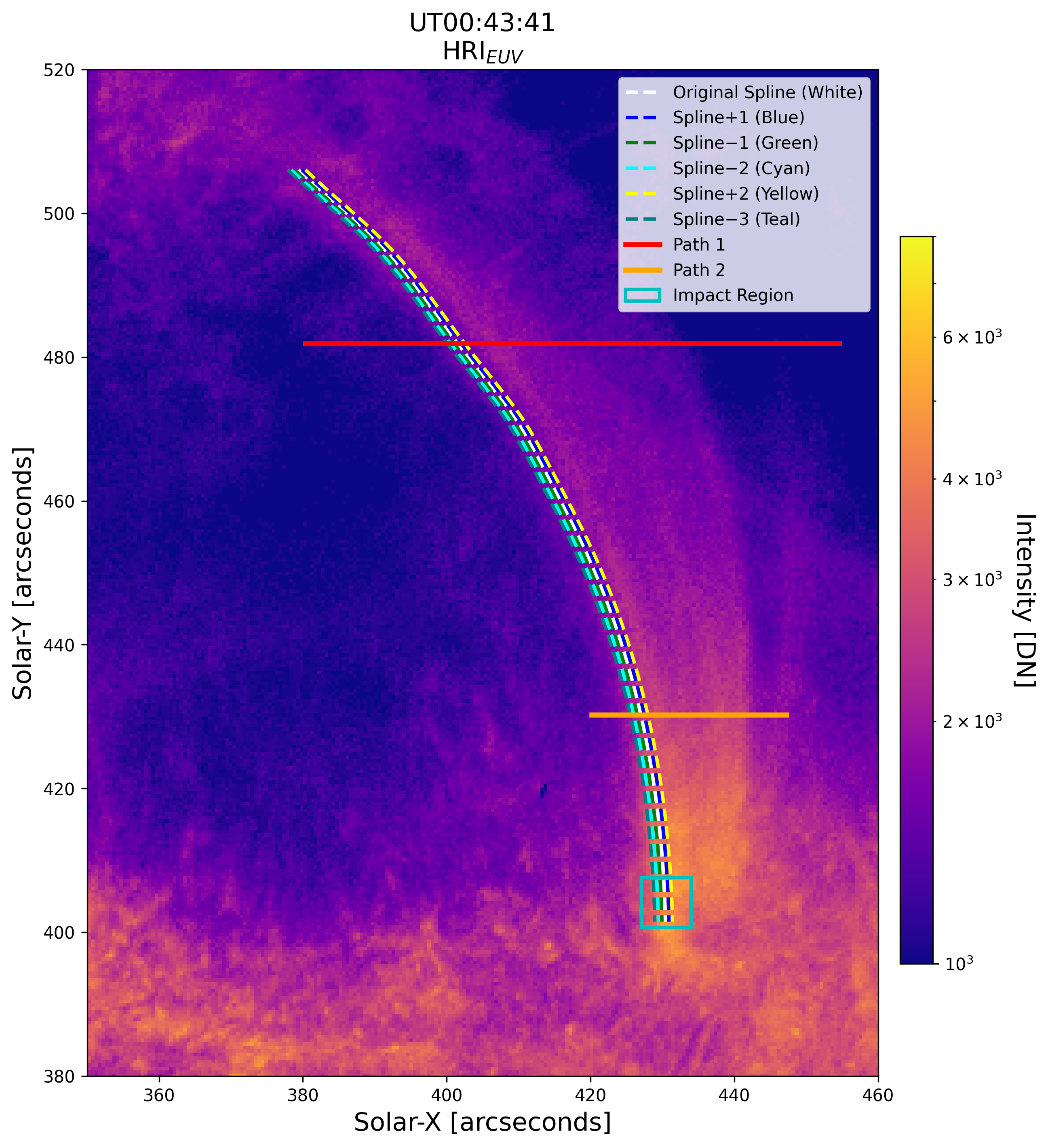}
    \caption{The coronal loop along which the coronal rain is seen to fall with \hrieuv. The `Original Spline' corresponds to the initial coordinates used to trace the rain trajectory. Spline$\pm x$ is a shift of the original spline by $x$ pixels along the $x$-axis, created to fully capture the extent of the rain shower. `Path 1' is used to assess the EUV variability in the loop, and `Path 2' is used to estimate the FWHM (Figure \ref{fig:hri_td_paths}). The `Impact Region' is where the rain is observed to impact the lower atmosphere. {We plot the intensity in logarithmic scale to make the coronal loop more visible.}}
    \label{fig:hri_sub_fov}
\end{figure}

\begin{figure}
    \centering
    \includegraphics[width=1.0\columnwidth]{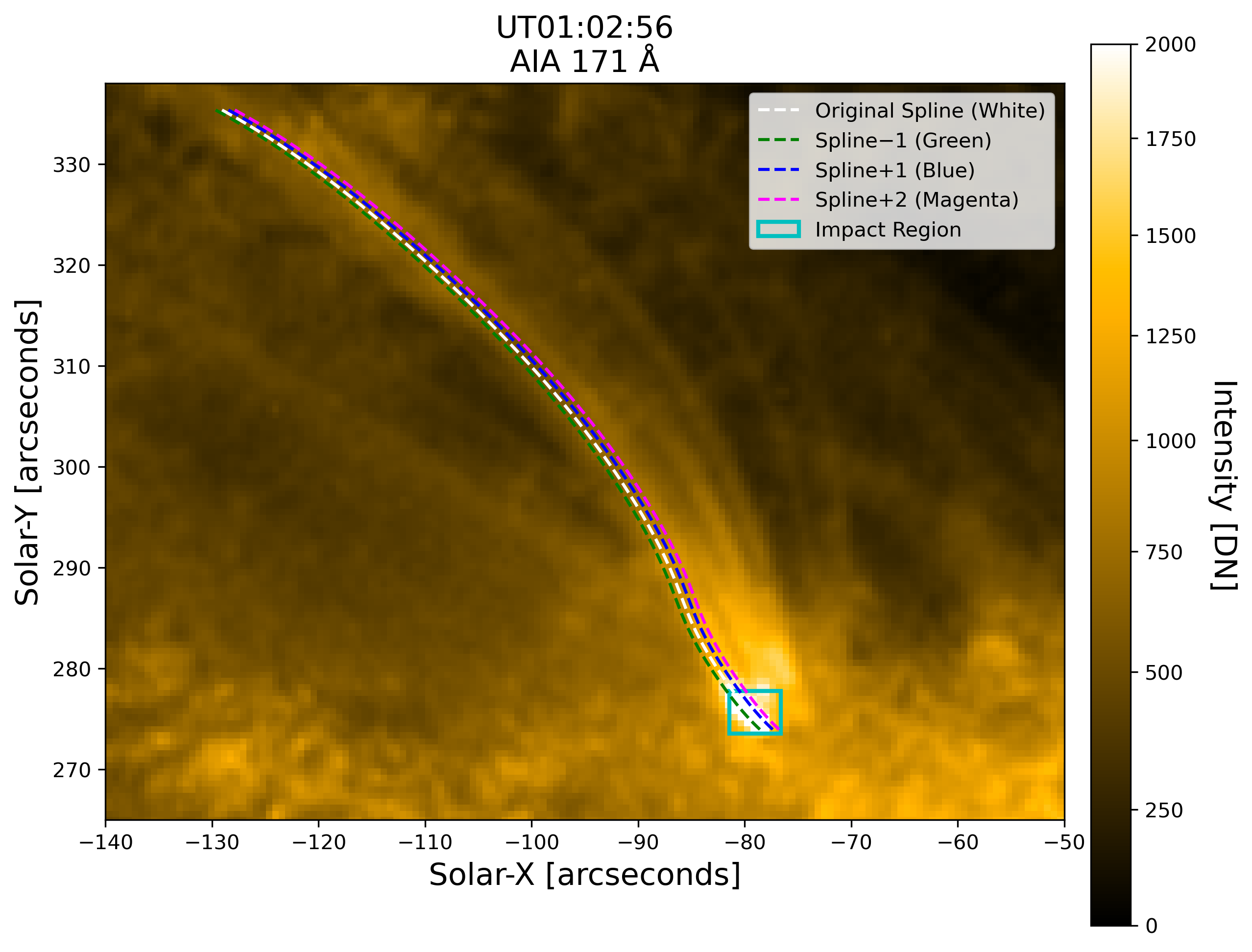}
    \caption{The coronal loop along which the coronal rain is seen to fall in AIA. Same as in Figure~\ref{fig:hri_sub_fov} but for AIA~171~\AA.}
    \label{fig:171_sub_fov}
\end{figure}

\begin{figure*}
    \centering
    \includegraphics[width=\textwidth]{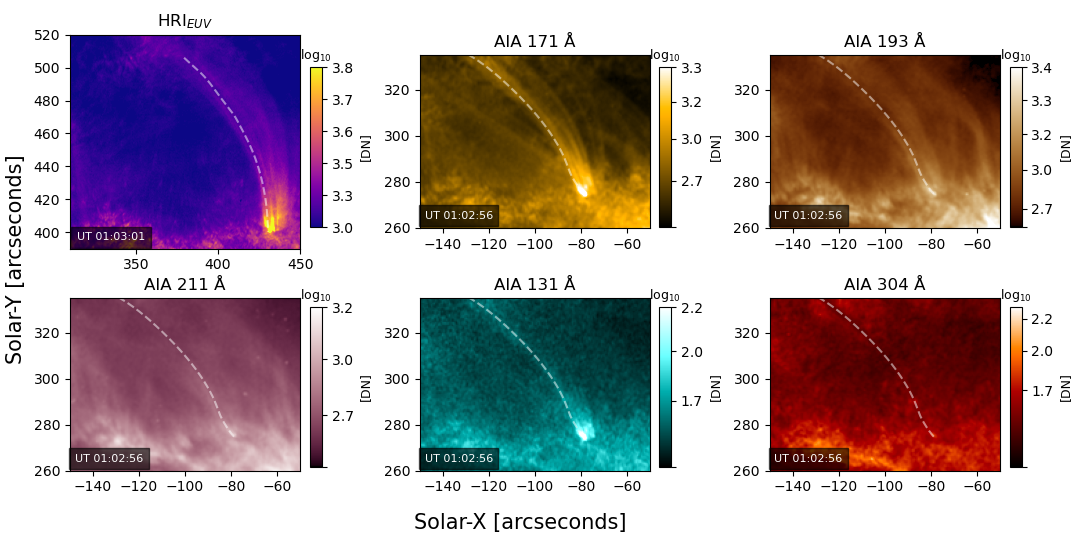}
    \caption{The coronal loop observed in the EUV channels, with the trajectory of the coronal rain clump shown by the white dashed curve at the time of impact of the rain clump with the lower atmosphere. See the online animation.}
    \label{fig:6_wavelengths}
\end{figure*}

To analyse the coronal rain event, we traced its trajectory with a series of 10 coordinates. Using those 10 coordinates, we generated a cubic spline interpolation, denoted as the `original spline' (in white) in Figures \ref{fig:171_sub_fov} and \ref{fig:hri_sub_fov}. {This is also shown in Fig.~\ref{fig:6_wavelengths} where the original spline is plotted for each wavelength.} As a single cubic spline did not capture the entirety of the coronal rain shower, we constructed multiple cubic splines and systematically shifted them along the $x$-axis relative to the original cubic spline as shown in Figures \ref{fig:hri_sub_fov} and \ref{fig:171_sub_fov}. The notation spline$\pm x$ is a shift of the original spline by $x$ pixels in the $x$-axis, created to capture the extent of the rain shower fully. Furthermore, the width of the rain clump is larger than one cubic spline, hence the need for multiple splines. This methodology was applied consistently in AIA, \hrieuv and SJI~1400~\AA.  

For AIA, the coronal rain path was first identified in 171~\AA\,  with the same cubic spline process previously explained. We then applied these same coordinates across all AIA wavelengths. The cubic splines depicted in Figures \ref{fig:171_sub_fov} and \ref{fig:hri_sub_fov} show the only region of the coronal loop where the coronal rain is observed. We then interpolated the intensity values along the cubic splines across all images to create the time-distance diagrams.

To determine the velocity of the descending rain and associated upward propagating features, we select the cubic spline along which the rain signatures are best seen and calculate the slope of the line in the time-distance diagram for that path. Using \hrieuv, the width of the coronal rain clump was assessed by taking a transverse cut across the coronal loop, denoted as path~2 in Figure \ref{fig:hri_sub_fov},  interpolating the intensity values along the path (path 2 in figure \ref{fig:hri_td_paths}) and calculating the Full Width at Half Maximum (FWHM). Additionally, `Path 1' was created to assess the EUV variations across the coronal loop.

We use the Differential Emission Measure (DEM) to estimate the mean temperature and number density of the hot, optically thin plasma. For a given temperature along the line-of-sight (LOS), the DEM at that temperature is defined as:
\begin{equation}
\label{eq:2.DEM}
\text{DEM}(T) dT = \int_{0}^{\infty} n_en_p(T) d\ell,
\end{equation}

where $n_e$ and $n_p$ are the electron and proton number densities, respectively, and $\ell$ denotes the distance along the LOS. The total Emission Measure (EM) is the integration of the DEM over the temperature range:

\begin{equation}
\label{eq:3.EM}
EM = \int_{T} \text{DEM}(T) dT,
\end{equation}
and the mean temperature can be estimated through the DEM-weighted temperature:

\begin{equation}
\label{eq:3.DEM_Weighted}
\langle T \rangle_{\text{DEM}} = \frac{\int_{T} \text{DEM}(T) T dT}{\text{EM}}.
\end{equation}

From equation \ref{eq:2.DEM}, we can therefore obtain the DEM-weighted electron number density ($\langle n_e \rangle_{\text{DEM}} $): 
\begin{equation}
\label{eq:3.ne}
\langle n_e \rangle_{\text{DEM}} = \sqrt{\frac{1.2 \int_{T} \text{DEM}(T) dT}{\ell}}.
\end{equation}
The factor 1.2 comes from the assumption of a highly ionized plasma with $10\% $ helium.

The DEM is calculated in IDL using the CHIANTI (version 10.1 developed by \citet{dere2023chianti}) atomic database, first introduced by \citet{dere1997chianti}. The method used is the regularization method discussed in \citet{hannah2012differential}. We selected 16 temperature bins, ranging from $\log_{10}T = 5.5$ to $\log_{10}T = 7.05$ with a bin size of 0.1 in logarithmic space.

\section{Results}\label{sec:results}

\subsection{Loop Properties}\label{sec:Loop properties}

\begin{figure}
    \centering
    \includegraphics[width=1.0\columnwidth]{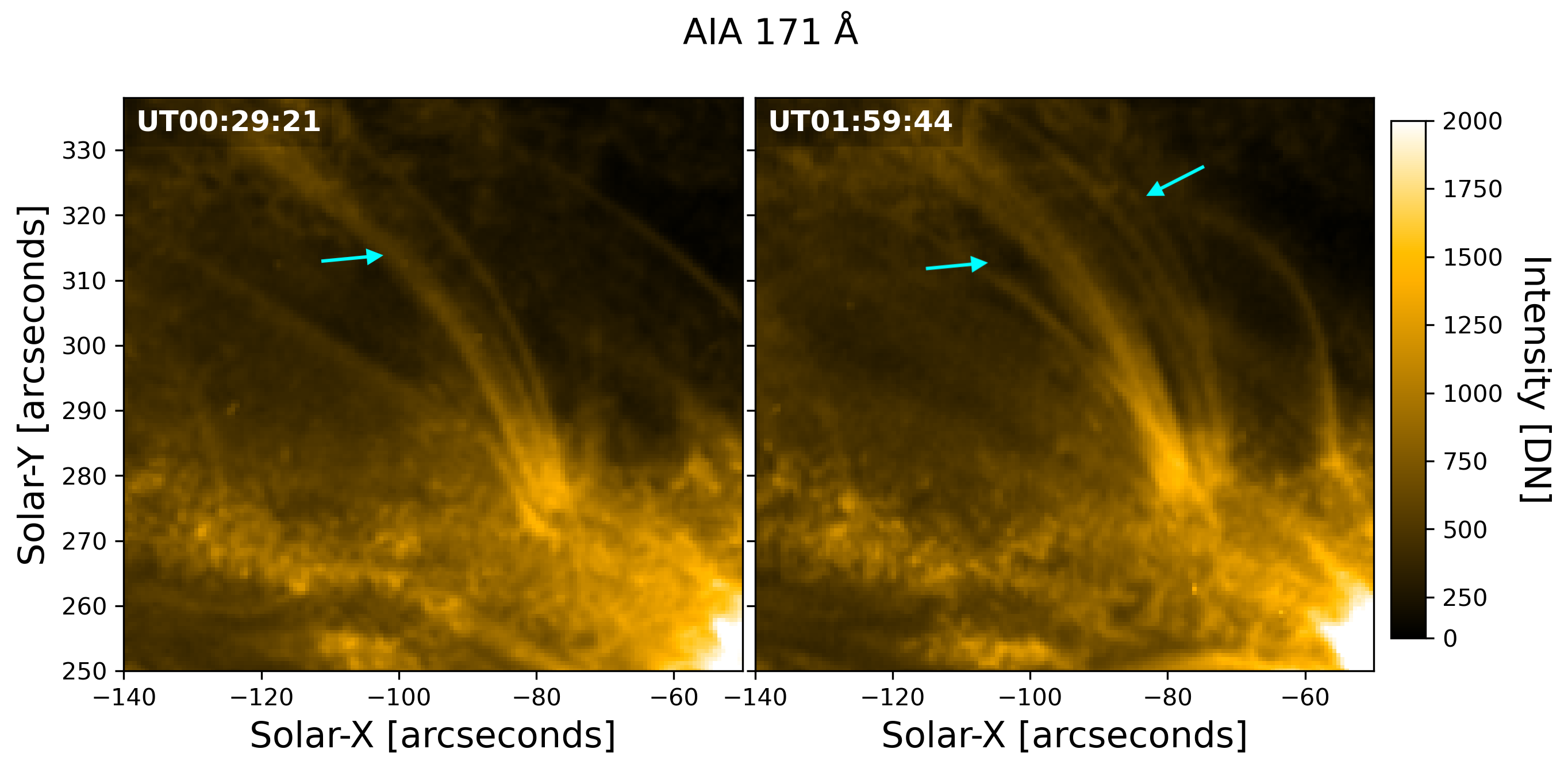}
    \caption{Sub-FOV of AIA 171 \AA, for the first (left) and last (right) snapshots of the observation. The cyan arrows point to coronal strands.}
    \label{fig:171_first_last}
\end{figure}

The complementary observations from \hrieuv, AIA and IRIS allow for a thorough analysis from two vantage points of the structure and morphology of the coronal loop affected by the coronal rain event. Although data from IRIS only captures the lower part of the loop, it allows us to pinpoint the loop's footpoint within the chromosphere and analyse the effects of coronal rain at lower atmospheric layers.

\subsubsection{Loop Footpoint}\label{sec:Loop footpoint}

As mentioned in section~\ref{sec:observations}, both footpoints appear to be anchored in moss shown by the green arrows in Figure \ref{fig:hri_full_fov}. We observe flash-like EUV brightenings of various sizes at and around the footpoint (see online animation). Multiple small-scale EUV brightenings (only a few pixel wide) appear with typical durations of 50~s, reminiscent of those reported in \citet{Berghmans_2021AA...656L...4B} at the smaller scales. In addition to these, brightenings on the scale of a Mm or more appear as the rain falls, which we describe in detail in section~\ref{sec:Rain Impact}. Periodic disturbances can also be seen propagating upward along the loop, but only very faintly.

\subsubsection{Loop Structure }\label{sec:Loop Structure}

\begin{figure}
    \centering
    \includegraphics[width=1.0\columnwidth]{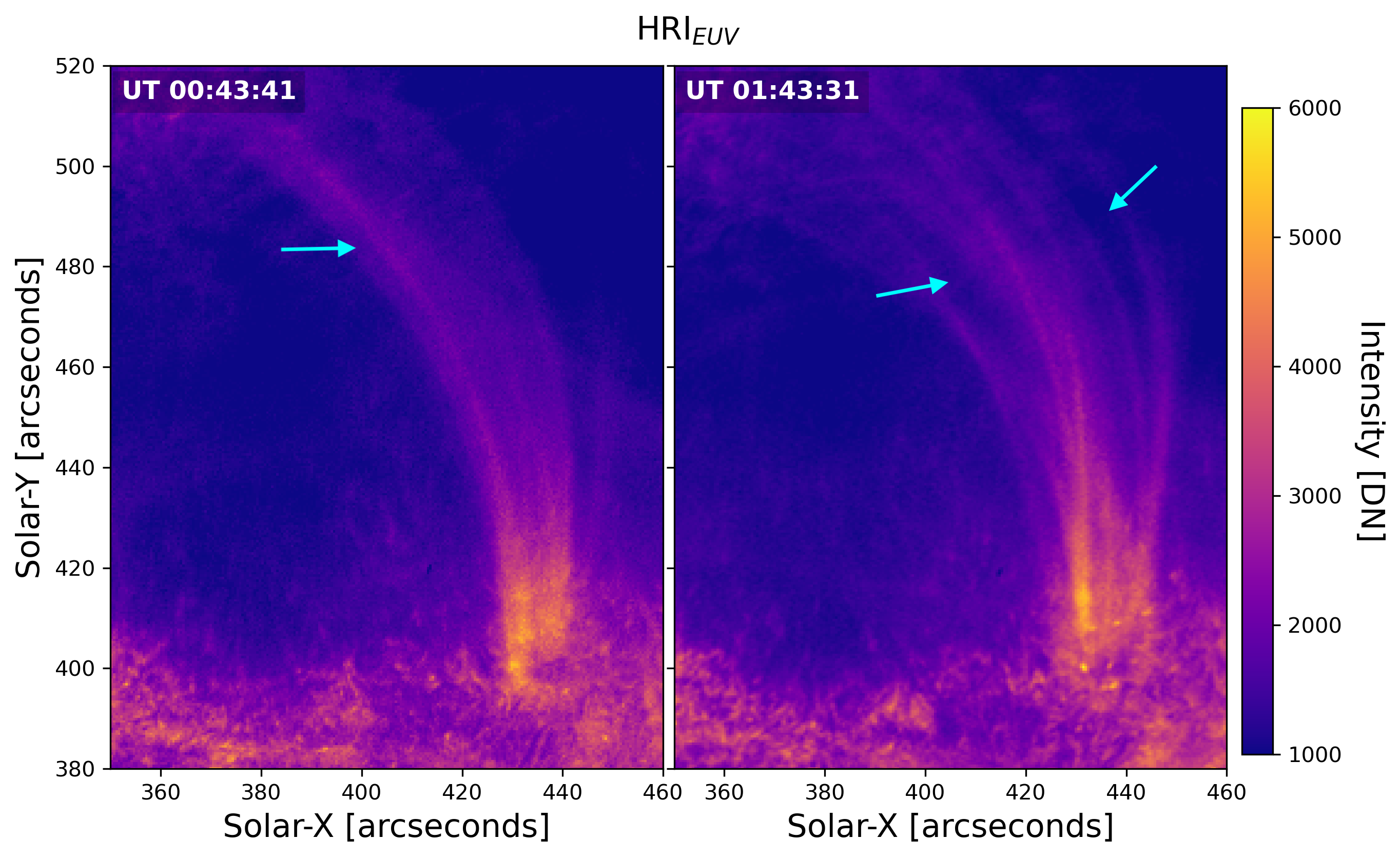}
    \caption{Same as in Figure~\ref{fig:171_first_last} but for \hrieuv.}
    \label{fig:hri_first_last}
\end{figure}

The estimated total length of the coronal loop observed with \hrieuv is approximately $160  \pm 3~$Mm. Throughout the observation, notable changes to the loop structure were observed. AIA~171~\AA\, and  \hrieuv show the clearest visibility of the loop, suggesting a temperature range of $0.6$ to $1$~MK on average. Figures \ref{fig:171_first_last} and  \ref{fig:hri_first_last} show the first and last images in AIA 171~\AA\, and \hrieuv, respectively. Initially, the width of the loop taken near the footpoint was $7.2\pm 0.3$ Mm and increased to $11.0\pm 0.3$~Mm by the end of the observation with \hrieuv. At the beginning of the observation in \hrieuv, the coronal strands are densely packed and exhibit almost uniform intensities, except for one notably brighter strand in the location where the coronal rain is observed to fall. By the end of the observation, new coronal strands appear, while others have disappeared, creating notable gaps between strands. Similar EUV variability with coronal loops is not uncommon and has been discussed in e.g. \citet{ugarte2006investigation}. Although AIA~171~\AA\, does not show coronal strands as compact as in \hrieuv, potentially due to the different LOS, slightly lower spatial resolution, it does exhibit similar morphological changes between the first and last images. 

At the beginning of the observation, AIA~94~\AA\, and 131~\AA\, dominated by \ion{Fe}{XVIII} (7 MK) and \ion{Fe}{XXI} (12 MK) respectively, depict minimal brightening and unclear visibility of the coronal strands. This unclear visibility suggests a lack of high-temperature plasma activity in these wavelength ranges. However, a brightening of a coronal strand, spanning $58  \pm  2$~Mm in length, on the western end of the loop, occurs approximately $15 - 20$~min before the coronal rain impacts, which is likely produced by the lower temperature emission in these passbands. 

As the rain falls, compression occurs ahead of the rain. This is composed on one hand of a small-scale brightening, better visible in \hrieuv, immediately ahead of the rain, known as the fireball effect (akin to meteor ablation), which has first been reported in \citet{antolin2023extreme}. On the other hand, a brightening on a larger scale is observed ahead of the rain a few minutes prior to the rain's impact (and culminates with the impact), and is discussed later on. After the impact, a new coronal strand near the loop's centre begins to brighten in \hrieuv. 

\begin{figure}
    \centering
    \includegraphics[width=1.0\columnwidth]{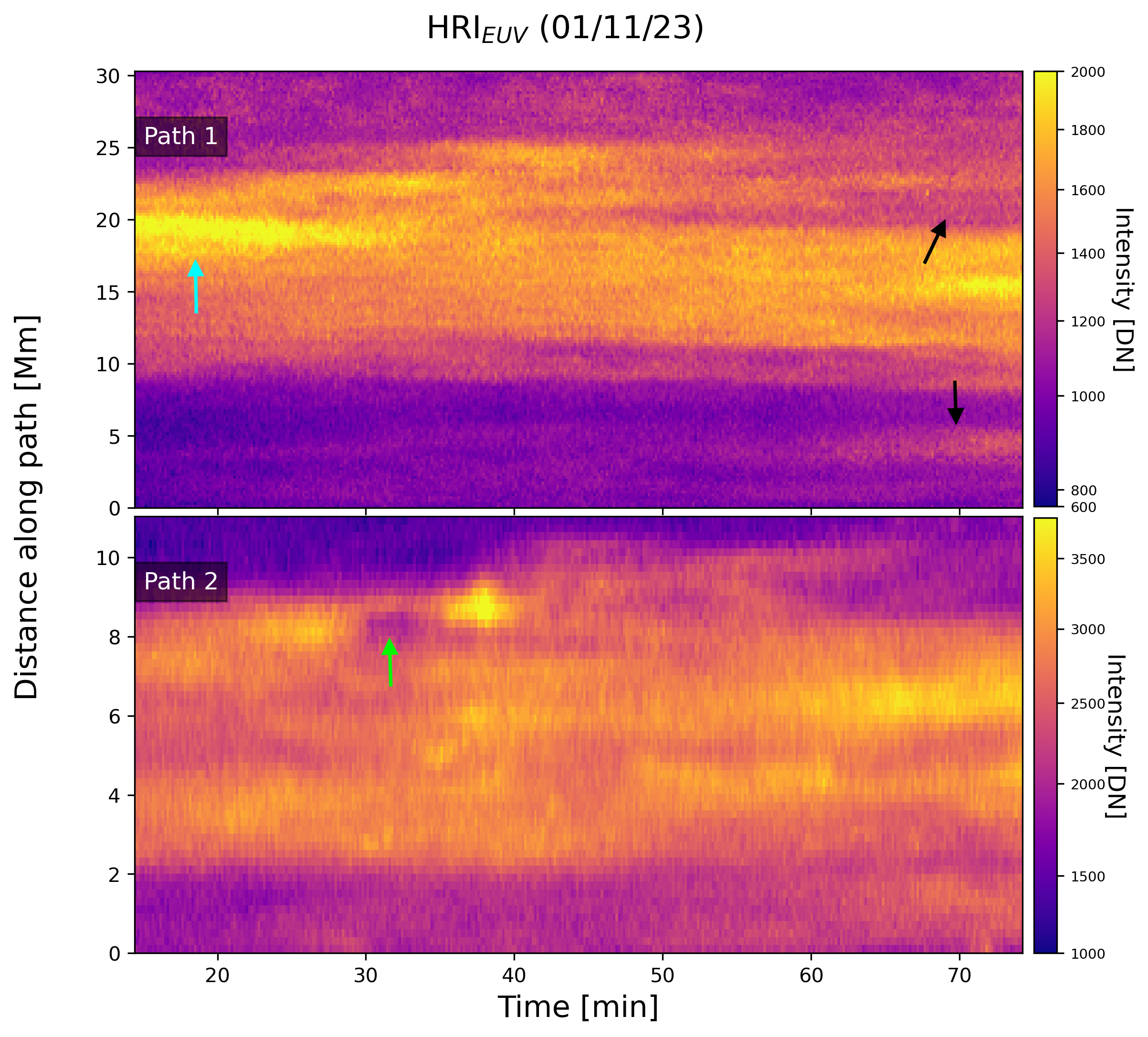}
    \caption{Time-distance diagrams of the two paths shown in Figure~\ref{fig:hri_sub_fov}. {Where 0~Mm corresponds to the right side of Path 1 and 2 in Fig.~\ref{fig:hri_sub_fov}.} The cyan arrow in `Path 1' points to a coronal strand that brightens prior to the rain falling. The black arrows highlight the EUV variability as bright strands appear to fade away and new strands emerge. The green arrow in `Path 2' points to an EUV absorption feature caused by the rain clump. {Note that it is preceded by a brightening feature, which corresponds to the compression analysed in section~\ref{sec:Compression by the rain}.}}
    \label{fig:hri_td_paths}
\end{figure}

Figure \ref{fig:hri_td_paths} shows the time-distance diagram along path 1 across the loop in \hrieuv (seen in Figure \ref{fig:hri_sub_fov}). We can see clear EUV variability in the coronal loop throughout the whole observation. Highlighted by the cyan arrow on `Path 1', we see a significant brightening of the same coronal strand where the rain is seen to fall, but at least 15~min prior to the rain passage (see online animation). During this time the strand slowly diminishes in intensity to match the average loop intensity by the time the rain crosses the path. Such brightening, also observed in multiple AIA passbands, is in agreement with thermal non-equilibrium, with parts of the loop catastrophically cooling down as the coronal rain forms.  Further down the loop, we took another path denoted as `Path 2' in Figure~\ref{fig:hri_td_paths}. We see a dim brightening corresponding to the fireball compression, followed by a large absorption feature (highlighted by the green arrow) matching to the time when the rain clump crosses that path, and therefore corresponding to the EUV absorption by the rain's cool material.

\subsection{Rain Properties}\label{sec:Rain Properties}

\subsubsection{Morphology and EUV variability}\label{sec:Morphology and EUV variability}

\begin{figure}
    \centering
    \includegraphics[width=1.0\columnwidth]{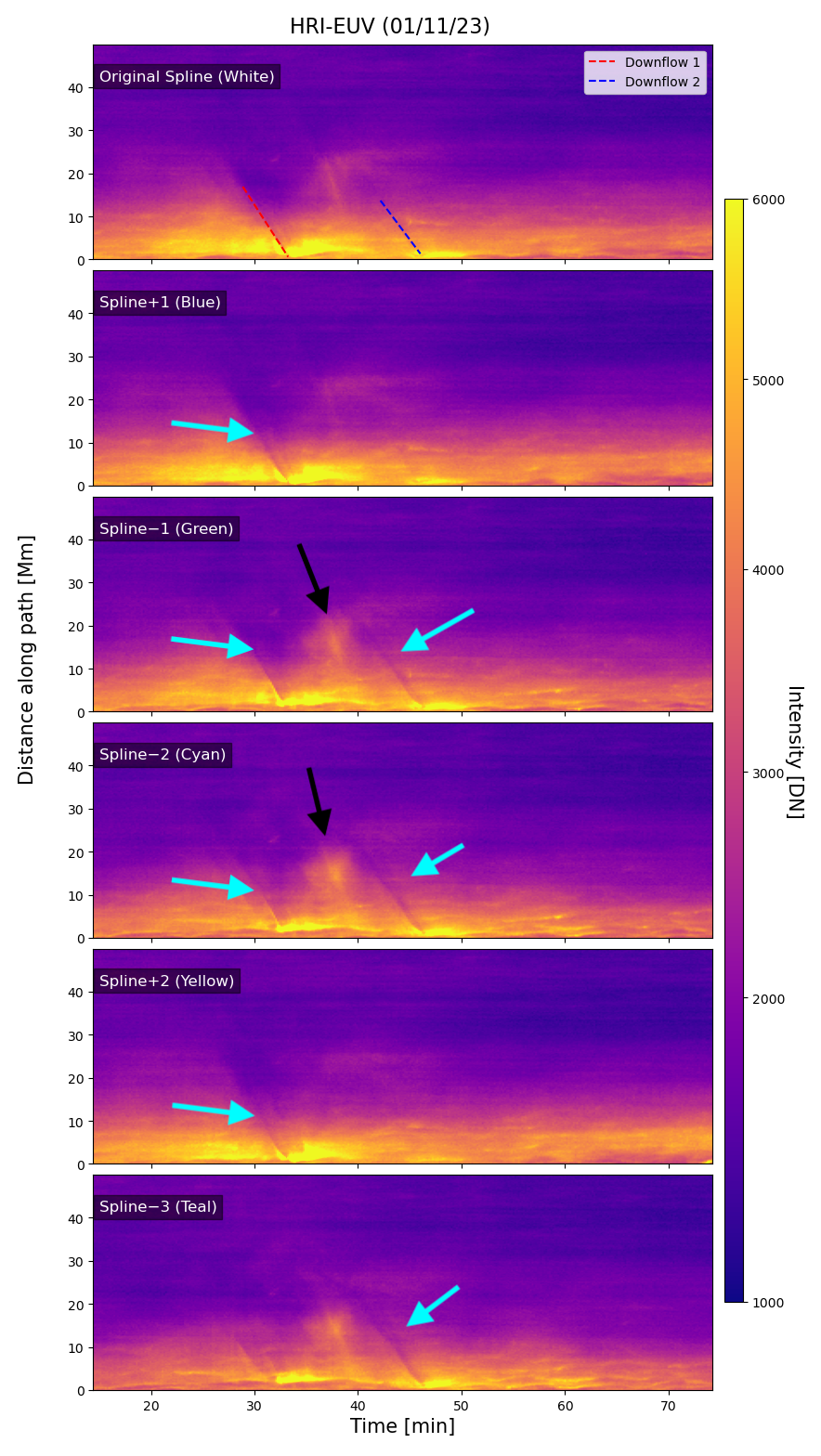}
    \caption{Time-distance diagrams extracted from the cubic splines in Figure~\ref{fig:hri_sub_fov}. Many rain clumps can be seen, notably those signaled by the cyan arrows, denoted as 'downflow 1', occurring between $t=28-35$~min (red dashed segment in the top panel), and 'downflow 2', between $t = 42-48$~min (blue dashed segment in the top panel). The black arrow indicates the presence of an upward flow (see Figure~\ref{fig:193_td_splines}).}
    \label{fig:hri_td_splines}
\end{figure}

\begin{figure}
    \centering
    \includegraphics[width=1.0\columnwidth]{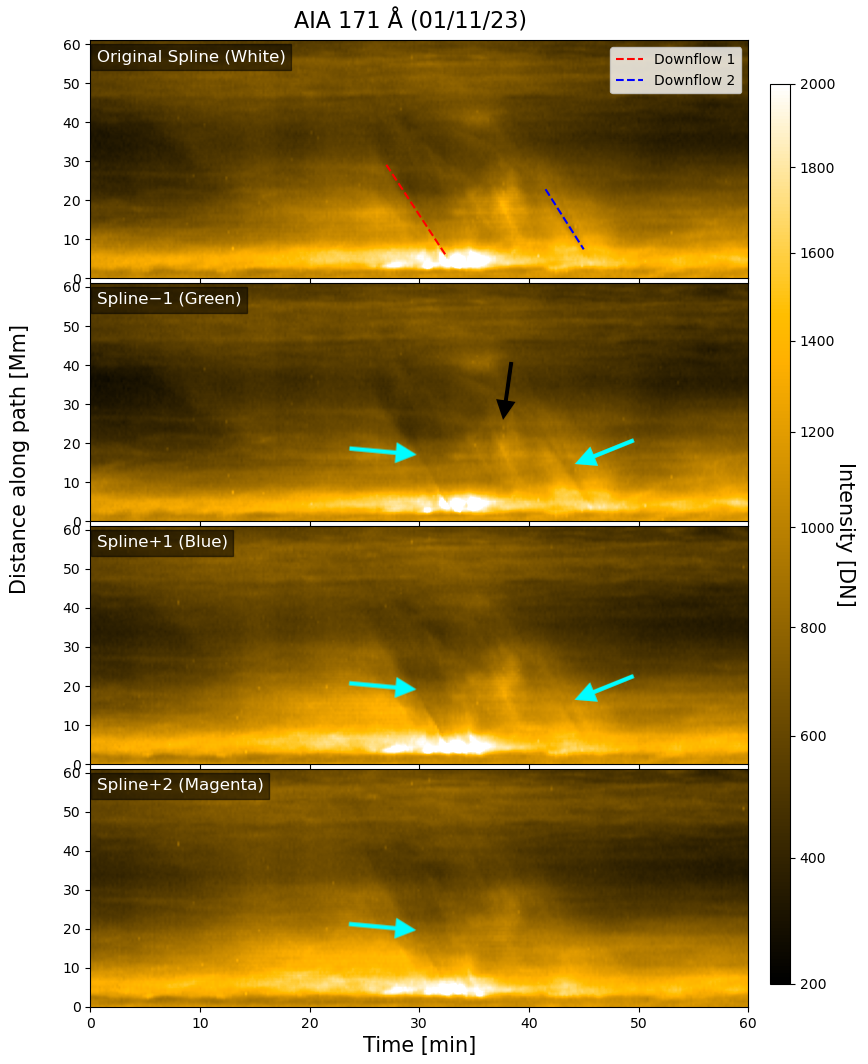}
    \caption{Similar to Figure~\ref{fig:hri_td_splines} but for the  splines plotted in Figure~\ref{fig:171_sub_fov} for AIA~171~\AA. downflows 1 and 2, identified in Figure~\ref{fig:hri_sub_fov}, can also be seen here clearly (respectively, red and blue dashed segments in top panel).}
    \label{fig:171_td_splines}
\end{figure}

\begin{figure}
    \centering
    \includegraphics[width=1.0\columnwidth]{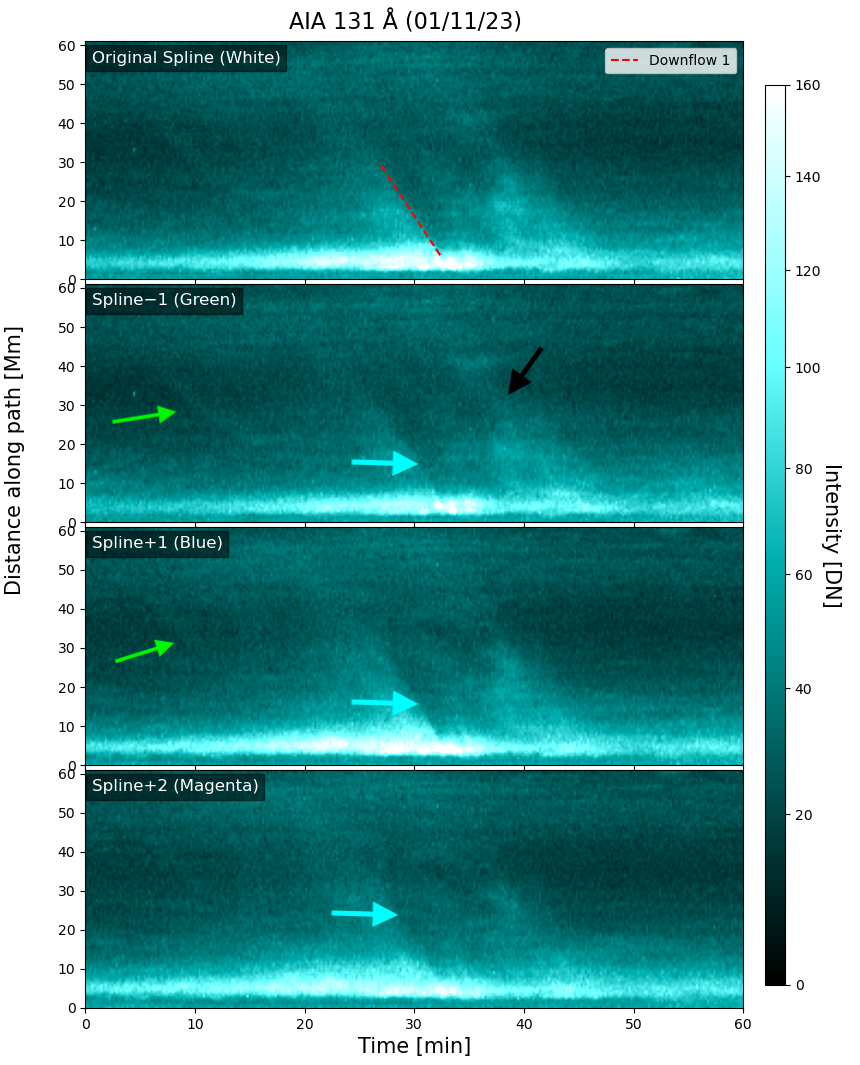}
    \caption{Similar to Figure~\ref{fig:hri_td_splines} but for AIA~131~\AA\, with the splines plotted in Figure~\ref{fig:171_sub_fov}. Note the presence of more downflow signatures prior to downflow~1 (indicated by the green arrows).}
    \label{fig:131_td_splines}
\end{figure}

In Figures \ref{fig:hri_td_splines} and \ref{fig:171_td_splines} we show the time-distance diagrams created from the cubic splines from Figures~\ref{fig:hri_sub_fov} and \ref{fig:171_sub_fov}, respectively. A downflow corresponding to coronal rain initiates roughly 25~min from the start of the \hrieuv observation and is observed to last for over 15~min in the time interval $t=27-35$ min. The downflow is characterised by a first clear rain clump, which we refer to as `downflow 1' (left cyan arrow in the figures).  downflow~1 can be identified in all the cubic splines shown, except the Teal spline (bottom panel in Figure~\ref{fig:hri_td_splines}), thereby indicating the extent of the rain shower.

To estimate the width of the clump, we use the transverse cut across the loop denoted as 'Path 2' in Figure~\ref{fig:hri_td_paths}, observed with \hrieuv. The aforementioned absorption profile produced by the clump is fitted with a Gaussian, and we use the Full Width at Half Maximum (FWHM) to define the clump's width. We estimate this width to be $0.55 \pm 0.03$ Mm. We note that this is not possible with AIA~171~\AA\, since the EUV absorption is not observed in that passband.

From the time-distance diagrams in Figure~\ref{fig:hri_td_splines}, we choose the white spline to calculate the speeds of the descending coronal rain based on visual clarity. The first clump denoted as downflow 1 in \hrieuv, descends at a constant projected speed of approximately $62.0\pm 10.0~$km~s$^{-1}$. The rain is seen in absorption, with emission ahead of the rain increasing in intensity as it falls, as expected from compression. Another clear downflowing feature, which we refer to as downflow 2, can be seen in Figure~\ref{fig:hri_td_splines} between $t=42-48$ min, which descends at a constant projected speed of $53.3\pm 10.0~$km~s$^{-1}$. No clear acceleration or deceleration of the clumps is observed by \hrieuv, suggesting a balance of forces (e.g. gas pressure and gravity) as the rain has reached terminal velocity, as discussed by \citet{oliver2014dynamics}. The rain is seen to descend an approximate distance of 60~Mm but is revealed more clearly in the last $20-30$~Mm in the time-distance diagram. 

As discussed in \ref{sec:methodology}, we first analysed the trajectory of the coronal rain event in AIA~171~\AA\, and used the same cubic spline coordinates in all other AIA wavelengths. The four cubic splines shown in Figure~\ref{fig:171_sub_fov} fully capture the extent of the rain shower from AIA's perspective. Using these splines we also show in Figures~\ref{fig:171_td_splines} and \ref{fig:131_td_splines} the time-distance diagrams for AIA~171~\AA\, and 131~\AA, respectively, which reveal more details about the rain shower. Indeed, these time-distance diagrams show signatures of more downflows prior to downflow~1, suggesting that the rain shower may start at $t\approx5~$min from the start of the observation and may last for roughly 35~min. As with \hrieuv, we also note the presence of the compression ahead of the rain in the last few Mm, as well as signatures of the `fireball' effect immediately beneath the rain.

Downflow~1 can be seen in 131~\AA, 171~\AA, 193~\AA, and 211~\AA, but not  AIA~94~\AA\, and 335~\AA. Across these wavelengths, downflows~1 and 2, fall with the same constant projected speeds of $71.2\pm 10.0~$km~s$^{-1}$ and $73.6\pm 10.0$~km~s$^{-1}$ respectively, with no signatures of acceleration or deceleration. The constancy in the speeds indicates that temperature does not affect the observed speeds, as is expected from coronal rain.

\begin{figure}
    \centering
    \includegraphics[width=1.0\columnwidth]{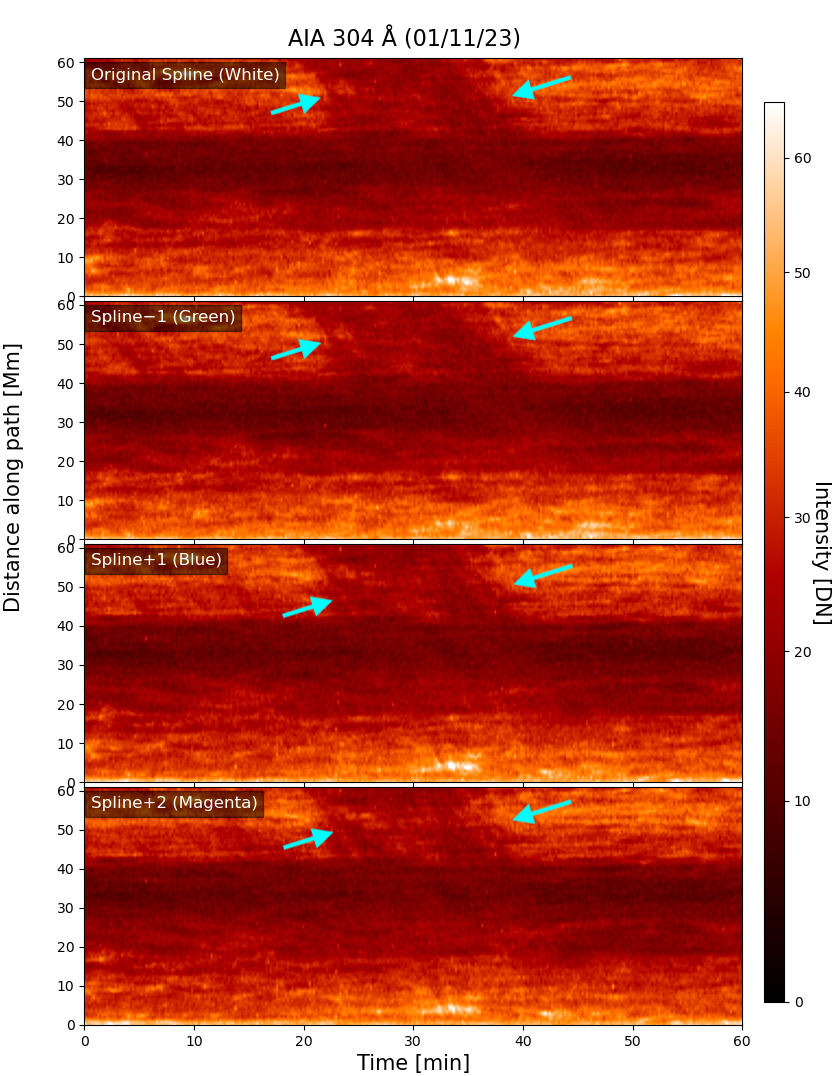}
    \caption{Similar to Figure~\ref{fig:171_td_splines} but for AIA~304~\AA\, using the splines plotted in Figure~\ref{fig:171_sub_fov}. The left cyan arrows point to a large absorption feature observed near the apex of the loop, which corresponds to the clump formation occurring between $t=22-35$~min leading to downflow~1. The right cyan arrow indicates downflow~2. Note the presence of further downflow signatures prior to downflow~1, also observed in Figure~\ref{fig:131_td_splines} with AIA~131~\AA.}
    \label{fig:304_td_splines}
\end{figure}

\begin{figure}
    \centering
    \includegraphics[width=1.0\columnwidth]{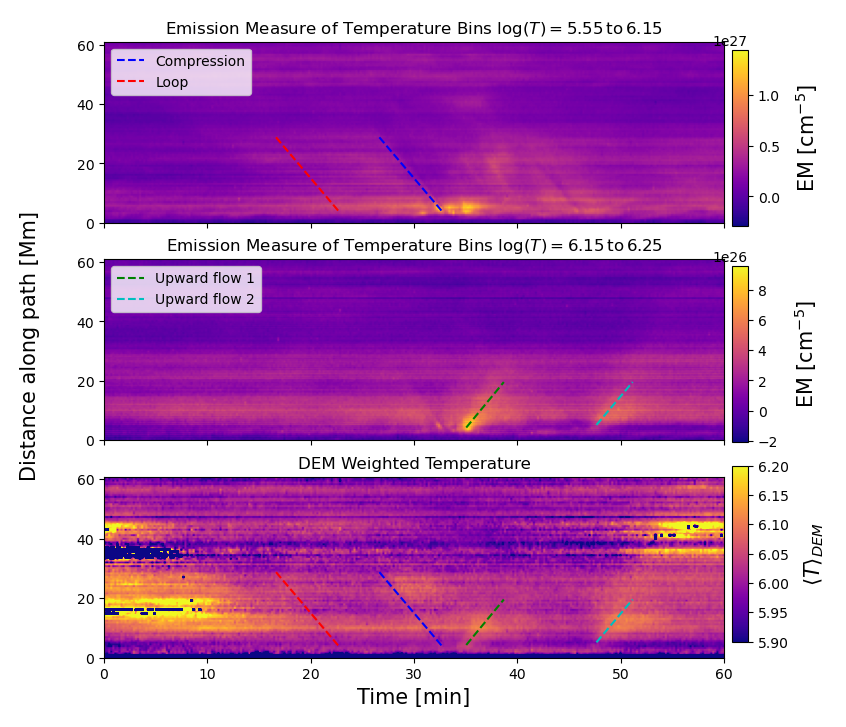}
    \caption{The top panel shows the total emission measure (EM) in the temperature range $\log(T)= 5.5 - 6.15$. The blue dashed line follows the compression by the rain. The red dashed line represents a parallel in time and space to the blue line, prior to the main rain event (downflow~1). Only temperature bins corresponding to $\log(T)= 6.15$ and 6.25 depicted the rebound flows. We denote `Rebound 1' in green, and `Rebound 2' in cyan in the middle panel. The bottom panel shows the DEM weighted temperature, where the loop, compression (due to downflow 1) and rebound flow lines are plotted.} 
    \label{fig:em_dem-weighted}
\end{figure}

In Figure~\ref{fig:304_td_splines} we show the time-distance diagram for AIA~304~\AA, which is adept to cooler temperatures of $\log T\approx5$ characteristic of the low transition region. We can see two large absorption features at $50-60$~Mm, matching the expected starting times of downflows 1 and 2 seen in the hotter wavelengths lower down the loop. These clumps can be seen throughout the fall, but with more or less clarity depending on the brightness of the background (the clumps can either absorb or emit stronger than the background). 

AIA~131~\AA, 171~\AA\, and \hrieuv, presented a clear EUV enhancement likely due to compression ahead of the descending rain, followed by absorption. In contrast, AIA~193~\AA\, and 211~\AA\, primarily showed a large absorption feature without the EUV enhancement ahead. This suggests that the compression due to descending coronal rain is in a temperature range of $0.4 - 1$~MK. This is confirmed by the DEM analysis shown in Figure \ref{fig:em_dem-weighted}, which shows significant emission only in the temperature range of $\log(T)= 5.5 - 6.15$.

By analysing the absorption created by the rain in Figure~\ref{fig:hri_td_splines} we estimated the length of downflow 1 to be  $1.1 \pm 0.3$ Mm. This length could be much longer \citep[as seen in previous rain observations, e.g.][]{antolin2012observing}, but not producing enough absorption to be observed.  However, we believe that downflow 1 is a collection of smaller, unresolved clumps (as seen in AIA~304~\AA) and to obtain a more accurate length, we {infer it} by multiplying the total velocity over the time between downflow 1 and the beginning of the upward flow (equal to {$120 \pm 30$ s}, which matches the time interval where most of the enhanced brightening is observed in the impact region). This leads to a length of  {$8.7 \pm 0.3$ Mm}. 

{We note that along all of the splines plotted we can detect traces of the rain falling in the corresponding time-distance diagrams, indicating that the rain shower has a width at least equal to the transverse extent of the splines, which is 1~Mm.} 

\subsubsection{Total rain speed}\label{sec:speed}

Using the projected velocities of downflows~1 and 2 from the two LOS provided by \hrieuv and AIA in Figures \ref{fig:hri_td_splines} and \ref{fig:171_td_splines}, we calculated the total velocities with the following equations. First, the separation angle between AIA and \hrieuv is:
\begin{equation}
\theta_{\text{EA}} = \theta_{\text{EUI}} + \theta_{\text{AIA}},
\end{equation}
where $\theta_{\text{EUI}}$ and $\theta_{\text{AIA}}$ denote, respectively, the separation angles between the plane-of-the-sky (POS) of \hrieuv and the coronal rain path, and between the POS of AIA and the coronal rain path. The projected velocity of the coronal rain in the POS of \hrieuv, $v_{\text{EUI}}$ is related to the total velocity $v$ as:
\begin{equation}
v_{\text{EUI}} = v \cos(\theta_{\text{EUI}}).
\end{equation}
Similarly, the velocity of the rain in the AIA's POS is:
\begin{equation}
v_{\text{AIA}} = v \cos(\theta_{\text{AIA}}).
\label{eg:7}
\end{equation}
These equations can be combined to obtain the total velocity $v$:
\begin{equation}
\label{eq:1}
v = \pm \sqrt{\left(\frac{v_{\text{AIA}} - v_{\text{EUI}} \cos(\theta_{\text{EA}})}{\sin(\theta_{\text{EA}})}\right)^2 + v_{\text{EUI}}^2}.
\end{equation}

The separation angle between SolO and AIA was obtained from the Stereo Science Centre website (\href{https://stereo-ssc.nascom.nasa.gov/cgi-bin/make_where_gif}{online}), where \(\theta_{EA} = 0.35\, \text{rad}\). Taking the speeds of downflow~1  to be  $v_{\text{AIA}} = 71.2$~km~s$^{-1}$ and $v_{\text{EUI}} = 62.0$~km~s$^{-1}$, we get a total velocity of $v = 72.5$~km~s$^{-1}$. Similarly, for downflow 2 we have $v_{\text{AIA}} = 73.6 $~km~s$^{-1}$ and $v_{\text{EUI}} = 53.3$~km~s$^{-1}$, we get a velocity of $v = 86.4$~km~s$^{-1}$.

\begin{figure}
    \centering
    \includegraphics[width=1.0\columnwidth]{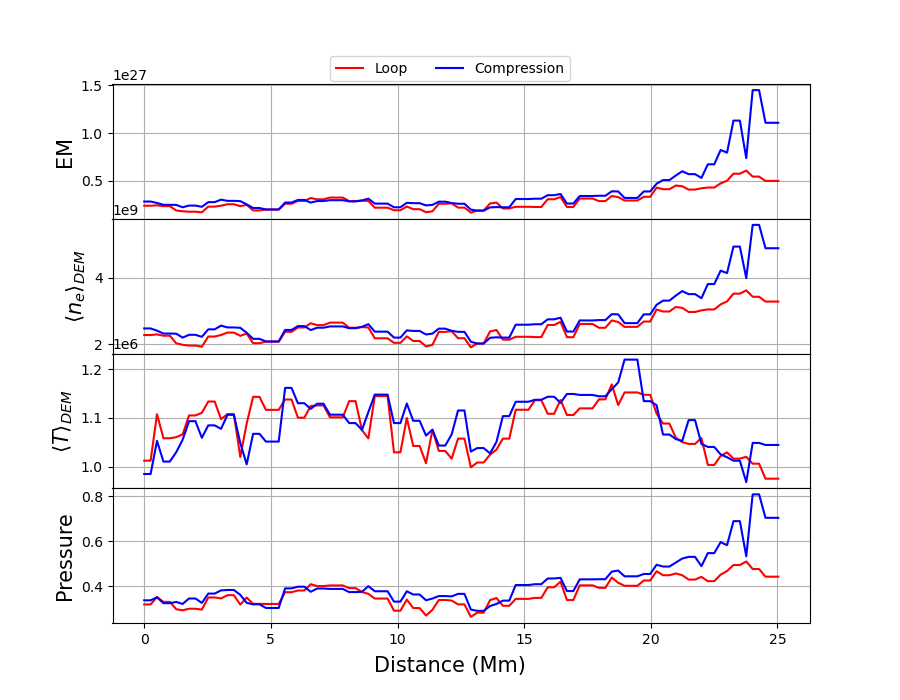}
    \caption{Thermodynamic quantities along the red (denoted as `loop' conditions) and blue (corresponding to compression) lines in Figure \ref{fig:em_dem-weighted}. We show the emission measure (EM, top panel),  the electron number density (Eq.~\ref{eq:3.ne}, second from top), the DEM weighted temperature (Eq.~\ref{eq:3.DEM_Weighted}, third from top), and the pressure (bottom).}
    \label{fig:ne_en_dem_p}
\end{figure}

\subsection{Compression by the rain}\label{sec:Compression by the rain}

We now analyse the compression produced by the rain during the fall and show that the ratio of specific heats (and polytropic index) can be estimated from it. For this, we first calculate the electron number densities and DEM-weighted temperatures in the compressed plasma. Using the methods described in Section~\ref{sec:methodology} we calculate these quantities along the blue and red lines in Figure \ref{fig:em_dem-weighted}, which follow, respectively, the compression produced by the rain {(subindex $c$)} and the parallel to this at a time 10~min prior, when the rain compression is not observed (but the conditions in the loop are otherwise assumed to be very similar). We denote the latter conditions (from the red line) as the `loop' conditions (without rain compression, {with subindex $l$}). The result is shown in blue in Figure~\ref{fig:ne_en_dem_p}. 

{To calculate the electron number density we took $\ell$ in Equation \ref{eq:3.ne}, as the width of the observed compression. We obtained a compression width of  $0.557 \pm 0.01$ Mm observed in \hrieuv and $0.690 \pm 0.01$ Mm observed in AIA 171~\AA. The larger compression width in AIA is likely due to the instrument's lower spatial resolution. Therefore, we used for $\ell$ the width obtained in \hrieuv. From Figure~\ref{fig:ne_en_dem_p} we see that the compression is to a large extent isothermal. At maximum compression we have $n_{e,c}= (5.59 \pm 0.19) \times 10^{9}~\text{cm}^{-3}$,  $n_{e,l}= (3.43 \pm 0.16) \times 10^{9}~\text{cm}^{-3}$, $T_{c} = (1.04 \pm 0.05) \times 10^6~\text{K}$, and $T_{l} = (1.00 \pm 0.09) \times 10^6~\text{K}$. That is, the density increases by a factor of 1.6, while the temperature only changes by a factor of 1.04 (i.e. 40,000~K).} 
 
{To understand this it is important to note that the rain acts like a cork for the transfer of energy upwards via thermal conduction. However, it can still be radiated in-situ in the compressed region or transferred downwards to the transition region. We estimate the conduction timescale $\tau_{\text{cond}}$, the radiative timescale $\tau_{\text{rad}}$ and the advective timescale $\tau_{\text{adv}}$, which is the time to impact of the rain. We have:}
\begin{eqnarray}
\tau_{\text{cond}}&=&\frac{3}{2}\frac{n_{e,c}L^2k_B}{10^{-6}T_c^{5/2}}\approx 300~\text{s}\\
\tau_{\text{rad}}&=&\frac{3}{2}\frac{k_BT}{n_{e,c}\Lambda}\approx120-170\text{~s}\\
\tau_{\text{adv}}&=&\frac{L}{v}\approx70~\text{s},
\end{eqnarray}
{where we have taken $L=5000~$km as the length over which the compression is observed.}  

{We can see that the condition of adiabaticity $\tau_{\text{adv}}<\tau_{\text{cond}},\tau_{\text{rad}}$ holds to some extent. If so, we have:}
%Because of the relatively low efficiency of thermal conduction in the compressed region (the energy cannot be conducted upwards due to the presence of the rain acting like a cork), and a low loss through radiation due to the relatively small time interval, we can assume the compression to be adiabatic, which implies:

\begin{equation}
T_c n_{e,c}^{-\gamma + 1} = T_l n_{e,l}^{-\gamma + 1},
\end{equation}
%where $T_c$ and $T_l$ denote, respectively, the compression and loop temperatures obtained from the DEM-weighted temperature. Similarly, $n_{e,c}$ and $n_{e,l}$ denote the electron number densities of the compression and of the loop, respectively, obtained from equation \ref{eq:3.ne}. 
where $\gamma$ denotes the ratio of specific heats. We then have:
\begin{equation}
\label{eq:3.gamma}
\gamma = 1 - \frac{\ln\left(\frac{T_c}{T_l}\right)}{\ln\left(\frac{n_{e,l}}{n_{e,c}}\right)}.
\end{equation}

{Using the values at maximum compression stated previously, we obtain} $\gamma = (1.085 \pm 0.218)$, reflecting the isothermal behaviour during the compression seen in Figure~\ref{fig:ne_en_dem_p}. {However, it is possible that the density of the compressed region just underneath the rain is actually not properly resolved, and much denser, in which case the radiative losses could be the main sink of energy.}
%and the reionisation of the neutral plasma in the rain.

\subsection{Rain Impact on the Lower Atmosphere}\label{sec:Rain Impact}

\begin{figure}
    \centering
    \includegraphics[width=1.0\columnwidth]{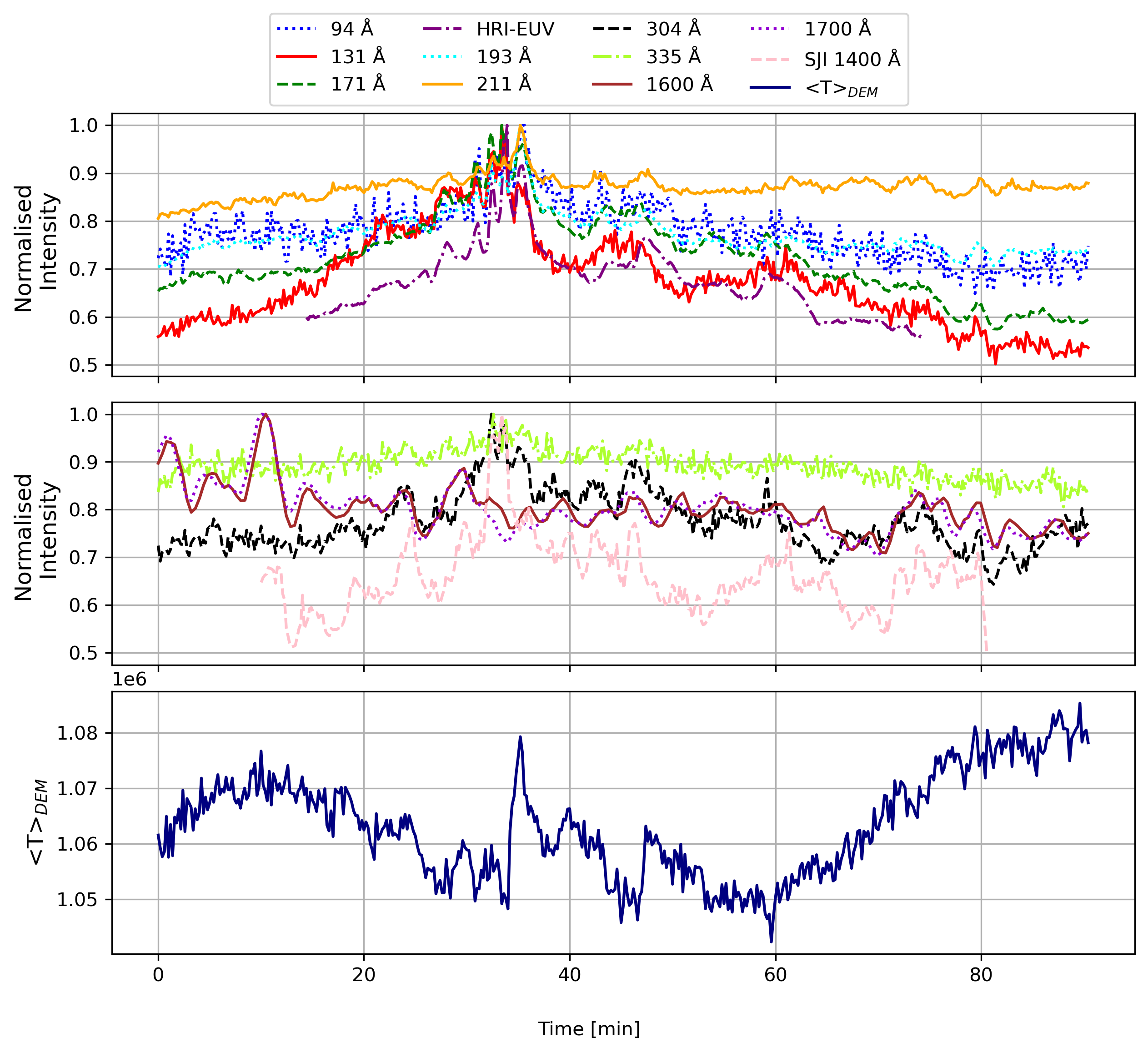}
    \caption{Multi-wavelength emission from the impact of coronal rain in the lower atmosphere (top and midlle panel). Co-alignment of UT times has been taken into account, and the time it takes light to travel to \hrieuv compared to AIA and SJI. We see a peak at around $t=30-35$ min in all intensities (except for AIA~1600~\AA), suggesting that the rain impacts deep into the lower transition region and compresses the plasma. The DEM-weighted temperature of the impact region is plotted in the lower panel.}
    \label{fig:impact_region}
\end{figure}

\begin{figure}
    \centering
    \includegraphics[width=1.0\columnwidth]{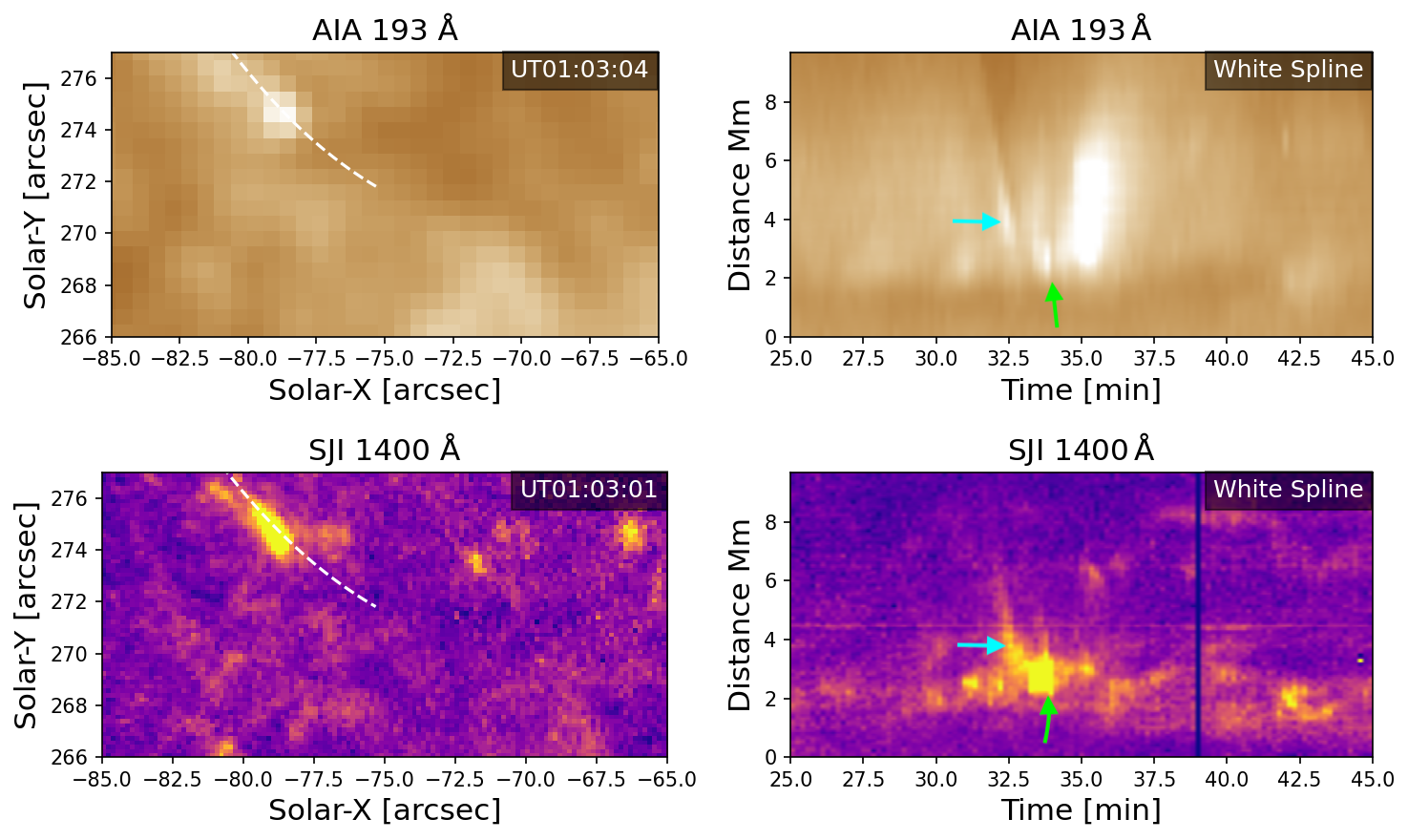}
    \caption{Rain impact detected by AIA and SJI. Left column: snapshot of AIA~193~\AA\, (top) and SJI~1400~\AA\, (bottom) immediately prior to impact, where the white cubic spline denotes the trajectory of the rain clump. Right column: the corresponding time-distance diagrams along the white cubic spline in AIA (top) and SJI (bottom). The time-distance diagrams clearly shows the {compression produced by downflow 1 (horizontal cyan arrow), followed by the impact (vertical} green arrow). The rebound flow can only be seen in the AIA~193~\AA\, time-distance diagram. See the online animation corresponding to this Figure. }
    \label{fig:iris_aia_snapshots_td}
\end{figure}

\begin{figure}
    \centering
    \includegraphics[width=1.0\columnwidth]{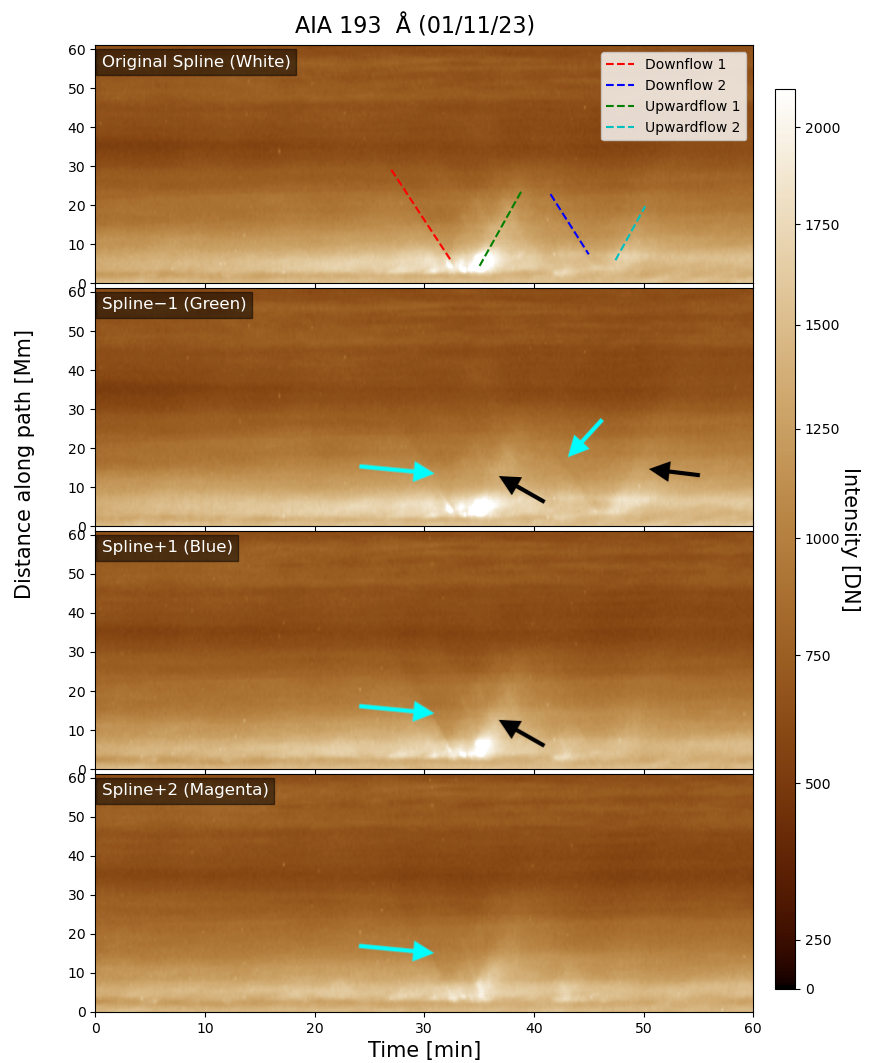}
    \caption{Similar to Figure~\ref{fig:hri_sub_fov} but for AIA~193~\AA\, with the splines plotted in Figure~\ref{fig:171_sub_fov}. The cyan arrows point to the descending clumps, while the black arrows point to the rebound flows. The speeds of the rebound flows are calculated using the green dashed lines (top panel).}
    \label{fig:193_td_splines}
\end{figure}

\begin{figure}
    \centering
    \includegraphics[width=1.0\columnwidth]{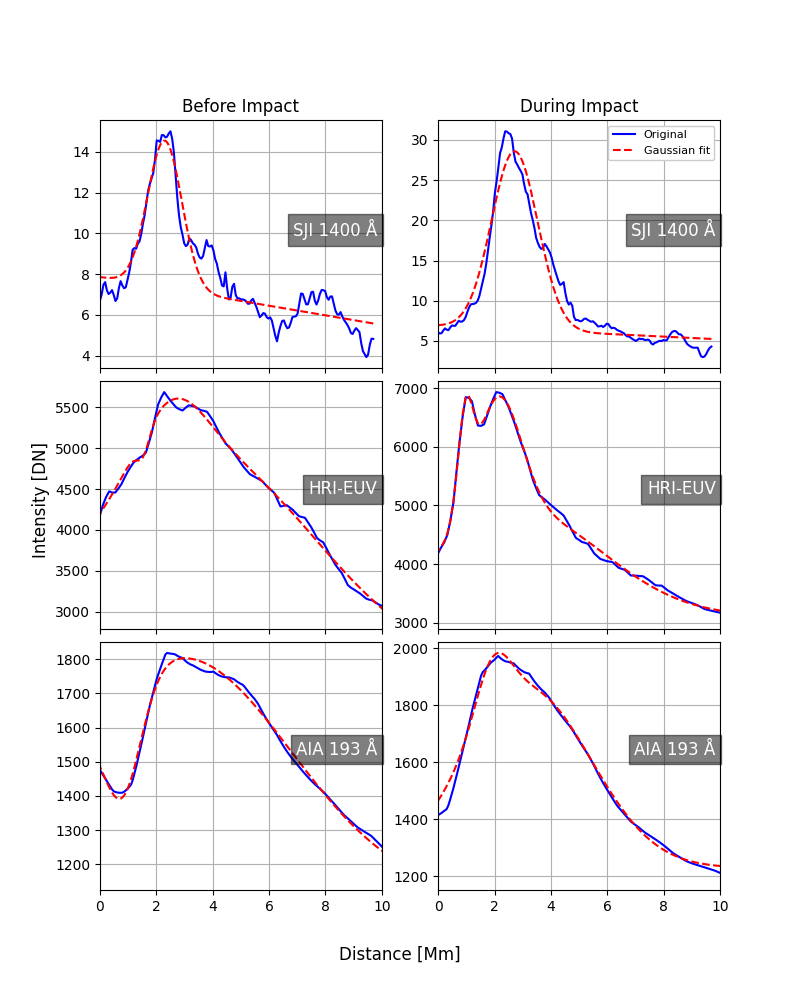}
    \caption{{The average intensities for SJI~1400~\AA\,(top), \hrieuv (middle) and AIA~193~\AA\ (bottom), before (left) and during impact (right), respectively. The averages are done with the white spline of their respective time-distance diagrams (Figure~\ref{fig:iris_aia_snapshots_td} and \ref{fig:hri_td_splines}) over the time interval $30.5 - 31.5$~min (before impact) and $33 - 34$~min (during impact).}}
    \label{fig:gaussian}
\end{figure}

To assess the impact of coronal rain on the lower atmosphere, we created a specific area at the footpoint of the loop known as the `Impact Region' in Figures~\ref{fig:hri_sub_fov} for \hrieuv and \ref{fig:171_sub_fov} for AIA. We verified that the selected region consistently covered the impact of the descending coronal rain across all the AIA wavelengths. The impact is also seen in SJI~1400~\AA\,  and AIA~304~\AA, suggesting that the rain reaches, at least, the lower layers of the transition region.

We summed the intensity values within these regions and plotted the light curves in Figure~\ref{fig:impact_region}. The light curves present a consistent pattern across multiple wavelengths, with an increase and simultaneous peak in intensity at around $t=30-35$~min, matching the time of impact of downflow~1, followed by a sharper decline. This is consistent across all the AIA EUV wavelengths, the \hrieuv and the SJI~1400~\AA. The increase and decrease trends are different across the wavelengths. The increase can be initially slow (over 30~min as seen in AIA~211~\AA), or fast (lasting only 5~min in AIA~94~\AA), and probably reflects the different processes at play, achieving different temperatures at different times. For instance, the intensity rise over $\approx20~$min seen in AIA~171\AA, 131~\AA\, and \hrieuv match the timescale over which compression is seen in the time-distance diagrams (and correspondingly, the intensity decline after the impact is sharper in these wavelengths). A second smaller intensity increase can be seen at $t\approx45-47~$min, and likely corresponds to downflow~2 since the timing matches the expected impact from the time-distance diagrams. 

AIA~1600~\AA\, and 1700~\AA\, exhibit no noticeable intensity variation around the times of the impacts. Hence, the rain does not seem to reach the chromosphere. 

We calculated the DEM weighted temperature for the impact region and plotted it in the bottom panel of Figure~\ref{fig:impact_region}. We notice very small variation on the order of $20-30~$kK over the entire observation, reflecting the isothermal behaviour of the compression ahead of the rain seen in Figure~\ref{fig:ne_en_dem_p}. The temperature actually exhibits a small decrease over the time of the impact of downflow~1, followed by a sharp peak of $\approx30~$kK matching the time of the rebound flow (discussed in the next section). A second smaller peak of $\approx$10~kK can be seen at $t\approx45-47$~min, which matches the time the rebound flow from downflow~2. This is then followed by a steady increase of 30~kK after $t=60~$min, which may suggest reheating of the loop and the start of a new TNE-TI cycle.  

In Figure~\ref{fig:iris_aia_snapshots_td} we take a close loop at the impact region as seen in AIA~193~\AA\, and SJI~1400~\AA. We note that the trajectory of the rain is slanted at the footpoint, indicating a similar topology for the magnetic field. Several small brightening events can be seen accompanying the impact of downflow~1, with an overall intensity enhancement over $2-3~$min (see also online animation). We notice the presence of the subsequent upflow following the rain impact in AIA~193~\AA\, but not in the IRIS channel, which we examine in the next section. We also notice that prior to the downflow~1 a more continuous (with little variability) intensity enhancement can be seen in both channels. This `steady' intensity enhancement can be seen in the full time-distance diagram of Figure~\ref{fig:193_td_splines} and in basically all the other channels. We note that the region in both channels at the base of the loop and underneath the impact region is dark.

\begin{figure}
    \centering
    \includegraphics[width=1.0\columnwidth]{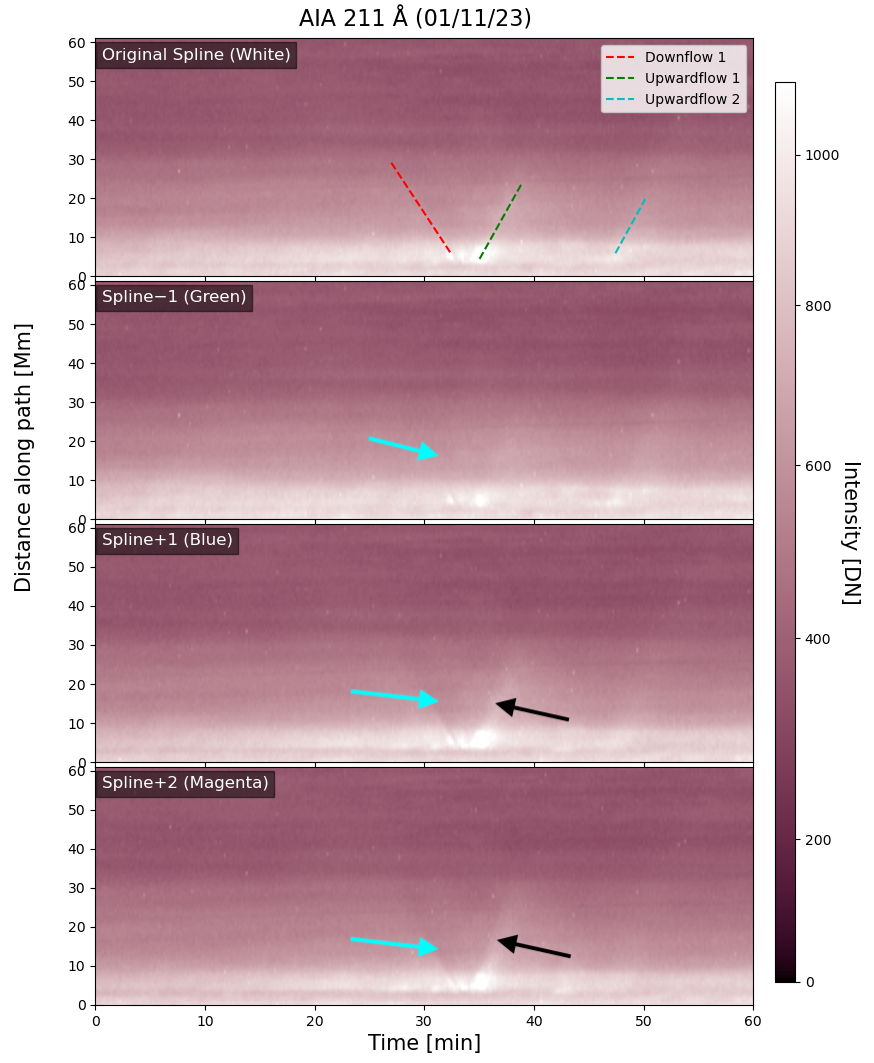}
    \caption{Similar to Figure~\ref{fig:hri_sub_fov} but for AIA~211~\AA\, with the splines plotted in Figure~\ref{fig:171_sub_fov}. The cyan arrow points to downflow~1, while the black arrow points to the rebound flow. The speed of the rebound flow is calculated using the green dashed line (top panel).}
    \label{fig:211_td_splines}
\end{figure}

To compare more closely the observed intensity variability across the {SJI~1400~\AA, \hrieuv and AIA~193~\AA\,} channels we plot in Figure~\ref{fig:gaussian} a time average over 1~min prior ($t=[30.5-31.5]~$min) and during the impact ($t=[33-34]~$min) for the intensities seen in the time-distance diagrams. {In this comparison it is worth to have in mind that while IRIS and AIA share the same LOS, Solar Orbiter does not. However, since the LOS is not so different (angle between both instruments is $\approx20^{\circ}$), we expect the changes produced by the different LOS to be small. This LOS difference introduces however an error in the co-alignment of features. Since we are not interested in an exact one-to-one correspondence at subarcsecond level (which is not possible given the different LOS), we address this uncertainty by taking the average over the time-distance diagrams for all the splines shown in Figures~\ref{fig:hri_sub_fov} and \ref{fig:171_sub_fov}, which fully cover the rain impact.}

{The panels in Figure~\ref{fig:gaussian} are organised in terms of temperature formation of the dominant spectral line in the passband, with lowest to largest from top to bottom.} Both the `steady' intensity enhancement and that from the impact can be seen as very similar large bumps in the intensity profiles (with an increased value for the impact). We measure the widths of these enhacements by fitting a single, double or {tripple Gaussian over the profile, as shown in the Figure. While the peak increases by a factor of two during the impact, the shape of the SJI profile is not greatly affected by the impact, and can be fitted with a single Gaussian} prior and during the impact, with a FWHM of 0.98~Mm and 1.29~Mm, respectively. {For \hrieuv, the peak intensity also increases but only by a factor of 1.3. The \hrieuv intensity profile is more structured around the peak and is best fitted by a tripple Gaussian. Prior and during impact, the narrowest (first) component has a FWHM of 0.53~Mm and 0.61~Mm, respectively. Similarly, the second component has a FWHM of 2.74~Mm and 1.89~Mm, respectively. Lastly, the third and broadest component has a FWHM of 9.67~Mm and 6.88~Mm, respectively. For AIA, the increase due to impact is smaller in amplitude, with a factor of 1.1.} Prior and during impact, AIA can be well fitted with a double Gaussian, with a narrow (first) component's FWHM of 1.18~Mm and 1.30~Mm, respectively. The broad (second) component of the fit has a FWHM of 4.76 Mm and 4.25~Mm, respectively. 

{It is very likely that the narrow profiles (in the case of SJI~1400~\AA\, the only profile) correspond to the transition region. The fact that the intensity increase is largest in this channel suggests that the rain clump reaches at least the low transition region, where it probably deposits most of its energy. We note that the width of the SJI~1400~\AA\, profile is similar to that of the second component of \hrieuv and the narrow component of AIA~193~\AA\ (prior and during impact). In contrast, the narrow width of \hrieuv is significantly smaller. These similarities and differences are likely due to differences in the response functions combined with differences in the spatial resolutions. For instance, SJI and \hrieuv have similar spatial resolution, but SJI~1400~\AA\, is sensitive to $10^{4.8}~$K plasma, while \hrieuv has a broader response with a peak at $10^6~$K and includes significant cooler contribution at $10^{5.5~}$K. Similarly, AIA has a factor of $\approx4$ lower spatial resolution, and AIA~193~\AA\, peaks at $10^{6.15}~$K but also includes cool contribution from $10^{5.4}~$K onwards. This is supported by the time-distance diagram of the emission measure of Fig.~\ref{fig:em_dem-weighted}, where we can see that the bright region near the footpoint for the emission measure bins $\log T=5.55-6.15$ is about half the size ($\lesssim5~$Mm) of the corresponding bright region for the emission measure bins $\log T = 6.15-6.25$ ($\lesssim10$~Mm). We also note that the AIA and SJI profiles show little change in their widths and peak location during the impact, which suggests a very narrow transition region less than 1.2~Mm in thickness. On the other hand, the \hrieuv intensity profile is more significantly affected. This reflects the higher sensitivity of the upper transition region to energy injection at lower heights,} in agreement with the results of \citet{Mandal_2023AA...678L...5M} where the studied dynamic fibrils with \hrieuv appear to show a bright front with a thickness of less than 1~Mm. 

%The fact that during impact, the AIA~193~\AA\, and SJI~1400\AA\, intensity enhancements (which share the same LOS) are observed at the same height in both channels suggests no change in the observed impact depth of the coronal rain clumps in SJI~1400~\AA\, compared to AIA~193~\AA\, further indicating that the entire transition region is affected by the impact. 
{Prior to the impact, we note that the broad (second and third components)} observed in \hrieuv, and AIA~193~\AA\, (and in many other AIA channels) are persistent, which could be interpreted as a footpoint localised steady heating signature, as expected by TNE-TI theory. {In this case, the FWHM of these components, which vary between 2~Mm and 10~Mm, could be a measure of the heating scale height.}
%To have a measure of the width of this transition we measure the thickness from 10\% the peak of the Gaussian fits to 10\% the background value on the left side of the Gaussian fits (this is marked by the green points in each Panel in Figure~\ref{fig:guassian}). In the case of AIA and \hrieuv, we do this with the Gaussian fit on the small bump, since this is the one located lower in the atmosphere. {We find values for SJI~1400~\AA\, of 1.05~Mm and 0.91~Mm before and during impact, respectively. Similarly, for \hrieuv we find 1.11~Mm and 1.23~Mm. Finally, for AIA~193~\AA\, we find 0.85~Mm and 0.96~Mm}. 

\subsection{Rebound Flow}\label{sec:rebound flow}

After the impact of downflow~1, a clear upward propagating feature originating from the impact region, termed `upward flow 1', was identified between $t=34-37$ min in AIA~193~\AA\, and AIA~211~\AA, respectively, in Figures \ref{fig:193_td_splines} and \ref{fig:211_td_splines}. A second upward propagating feature can be seen (best in AIA~193~\AA) between $t=44-48$ min (`upward flow 2'), following the impact of downflow~2.  Upward flow 1 has constant projected speed of $83.6\pm 10.0~$km~s$^{-1}$ in both AIA~193~\AA\ and 211~\AA. Similarly, upward flow 2 has a constant projected speed of $85.0\pm 10.0~$km~s$^{-1}$ in AIA~193~\AA. Both rebound flows occur almost instantly following the impact of the descending coronal rain. Hence, we interpret these propagating features as rebound flows because of the relatively low speeds compared to the sound speed.

The fact that we observe these upflows very clearly in AIA~193~\AA\, and 211~\AA\, indicates relatively hot temperatures reaching $1.25 - 2$~MK. This is supported by the DEM analysis in Figure~\ref{fig:em_dem-weighted}, which shows that only temperature bins corresponding to a temperature range of $\log(T)= 6.15$ to 6.25 revealed the rebound flow. Upon close inspection, other wavelengths such as AIA~94~\AA, 131~\AA, 171~\AA, and \hrieuv, also show signatures (but less distinct) of these upward flows denoted by the black arrows in the respective time-distance diagrams. The distance along the loop the upflows are seen to reach in both the AIA~193~\AA\, and 211~\AA\, is approximately $58 \pm 3$~Mm. 
$53.3\pm 10~$km~s$^{-1}$
Using the projected and total velocity of downflow 1 and rearranging equation \ref{eg:7}, we obtain \(\theta_{AIA} = 0.19\text{ radians}\). Using this, we can calculate the total velocity of upward flows 1 and 2 in Figure \ref{fig:193_td_splines}. Taking the projected velocity of upward flow 1 to be $83.6\, \text{km~s}^{-1}$, we get a total velocity of $85.2\, \text{km~s}^{-1}$. Taking the projected velocity of upward flow 2 to be $85.0\, \text{km~s}^{-1}$, we get a total velocity of $v = 86.5\, \text{km~s}^{-1}$.

Using AIA~193\, we estimated the width of upward flow 1 using the FWHM to be $0.86\pm 0.007$~Mm. We further estimated the length using Figure \ref{fig:193_td_splines} to be $7\pm 0.5$~Mm. Upward flow 2 is not seen as clearly in order to accurately calculate the width.

\subsection{Energy Estimate}\label{sec:Energy Estimate}

We can now estimate the kinetic energy of the rain clump, $K_e = \frac{1}{2}mv^2$, where $m$ and $v$ denote, respectively, the rain clump's mass and total velocity. For this, we must first calculate  the electron number density of the clump, which is determined assuming pressure balance between the rain clump and the compression ahead of the rain. We can also assume that the temperature of the clump is on the order of $10^4$~K (and probably less), given that it produces EUV absorption (due to neutral hydrogen and helium, and singly ionised helium). 

We obtain an electron number density for the clump of $(5.86 \pm 0.37) \times10^{11}$~cm$^{-3}$, which is in line with usually reported density values for coronal rain \citep{antolin2022multi}. Based on the rain volume (assuming cylindrical geometry), and using the morphological properties {(width of $0.55\pm0.03$~Mm, length of $8.7\pm0.3$~Mm)} mentioned in section~\ref{sec:Morphology and EUV variability} and observed total descending speed {($=72.5~$km~s$^{-1}$)} calculated in section~\ref{sec:speed}, we obtain a kinetic energy for the rain clump of {$(6.18 \pm 1.86)\times 10^{25}$}~erg. {Using the temperature of 10,000~K for the rain clump we calculate the thermal energy, given by $U = \frac{3}{2} k_B V T n_{\text{tot}}$, where $V$, $T$ and $n_{\text{tot}}$ denote, respectively, the volume, temperature and number density of the rain clump. We find a value of $(4.79\pm0.86)\times10^{24}$~erg.}

We also calculate the thermal and kinetic energy of the rebound flow to analyse how much energy from the rain clump is transferred to it. Using the values for the upward flow found in section~\ref{sec:rebound flow} we find thermal and kinetic energies of $(1.01 \pm 0.25)\times 10^{24}$~erg and $(9.24 \pm 0.8)\times 10^{24}$~erg, respectively. We notice a factor of {$\approx6.5$} difference between these values and the rain clump energy.

The rain impact will also likely generate MHD waves, and particularly slow modes. Such modes would propagate upward at the sound speed. Although they do not show up in the time-distance diagrams, it is useful to estimate their kinetic energies. These are well approximated by $E_w = \frac{1}{2} \rho  (\delta v)^2 c_s$, where $\delta v$ is the amplitude and $c_s$ is the sound speed. Since we do not observe the slow modes we can assume a maximum amplitude of 10\% the sound speed. We also assume the same volume as for the flow and obtain a wave kinetic energy value of $3.9\times10^{22}$~erg, which is negligible compared to the flow energy. 

We therefore find that {roughly} {$15\%$} of the clump's kinetic energy is transferred to the rebound flow. Since the impact is seen across all EUV and UV wavelengths (except AIA~1600 and 1700) we can deduce that most of the energy is radiated away in the transition region upon impact. For convenience we summarise all calculated values in Table~\ref{table:2}.

\begin{table*}
\centering
\renewcommand{\arraystretch}{1.2}
\begin{tabular}{lccc}
\hline
{Parameter}                 & {Downflow 1}                & {Upward Flow 1}               & {Compression} \\
\hline
Temperature $T$ (K)                & $10^{4}$                           & $(1.63\pm 0.07 ) \times 10^{6}$     & $(1.04 \pm 0.06)\times 10^{6}$\\
Electron density $n_e$ (cm$^{-3}$) & $(5.86 \pm 0.37)\times 10^{11}$    & $(3.50 \pm 0.06) \times 10^{9}$     & $(5.59 \pm 0.19) \times 10^{9}$ \\
Pressure (dyn cm$^{-2}$)           & ${1.55} \pm 0.08 $          & ${1.51} \pm 0.06$                     & $0.81 \pm 0.05$ \\
Velocity (km/s)                    & $72.5 \pm 10.0$                    & $85.2\pm 10.0$                     & -- \\
Width (km)                         & $549 \pm 27$                       & $862\pm 7$                         & ${557} \pm 1$ \\
Length (km)                        & $8702 \pm 300 $                    & $7000 \pm 500$                     & -- \\
Kinetic Energy (erg)               & {$(6.18 \pm 1.86)\times 10^{25}$}     & $(1.01 \pm 0.25)\times 10^{24}$    & -- \\
Thermal Energy (erg)               & $(4.79 \pm 0.86)\times 10^{24}$    & $(9.24 \pm 0.8)\times 10^{24}$    & -- \\
\hline
\end{tabular}
\caption{Physical quantities for downflow 1, upward flow 1, and the compression occurring ahead of the downflow 1.}
\label{table:2}
\end{table*}

\section{Discussion and Conclusion}\label{sec:Discussion and Conclusion}

This study provides a multi-wavelength observation of quiescent coronal rain and its effects on the lower and upper atmosphere with the help of instruments such as \hrieuv, AIA and IRIS/SJI. The coronal rain is observed on disc and is seen to fall onto a leading polarity region. The loop hosting the rain has both footpoints anchored in moss, where small and rapid EUV intensity enhancements are observed with \hrieuv, similar to those reported by \citet{Berghmans_2021AA...656L...4B}. We analysed the dynamics of the rain, its impact in the lower atmosphere and the morphological changes of the coronal loop that appear related to the rain. 

Throughout the entirety of the event, we observed strong EUV variability both during the rain shower (where downflows 1 and 2 are observed) and after its impact on the lower atmosphere. At the start of the observation, the coronal strands appeared densely packed and uniform in intensity, and by the end of the observation, we observed multiple strands disappearing with new ones appearing and notable gaps between strands. Such EUV variability is in agreement with that reported from loops in TNE-TI \citep{antolin2023extreme}, also in line with the generally observed variability in coronal loops \citep{ugarte2006investigation, antolin2022multi}. 

Although only two main clumps were very clearly detected (with EUV absorption signatures in \hrieuv), there appeared to be multiple smaller clumps falling, consistent with a large shower. This is supported by the streaks of lines prior to downflow 1 in \hrieuv and AIA (131~\AA, 171~\AA\, and 304~\AA) time-distance diagrams, and also during downflows 1 and 2 (observed in all time-distance diagrams). Overall, the rain shower appears to last about 35~min. The total length of the loop is approximately 160 ~Mm with the clump being observed to fall approximately 58~Mm. As depicted in the time-distance diagrams, only the last $20-40$~Mm reveals the clumps clearly. All EUV-UV channels, except from AIA 1600~\AA\, and 1700~\AA, detect the descending clump and the impact as depicted by figure \ref{fig:impact_region}. This supports coronal rain clumps being multi-thermal as discussed in \citet{antolin2015multi}. This is further supported by the DEM analysis in \ref{sec:Compression by the rain} where we detect the presence of coronal rain in temperature bins $0-7$, which have a temperature range of $\log(T)= 5.5 - 6.25$. 

The main coronal rain clump produced EUV absorption, with compression occurring ahead of the rain, particularly strong in the last minute prior to impact (up to a height of $\approx3$~Mm). The width of this compression is almost exactly the same as that of the rain width, reflecting the low plasma $\beta$ conditions and field-aligned motion of the plasma. Indeed, the 2D MHD numerical simulations by \citet{MartinezGomez_2020AA...634A..36M} show different compression and dynamics (with Kelvin-Helmholtz instability vortices at the wake of the rain) in the case of high plasma $\beta$. An additional brightening due to compression and heating immediately below the condensations is also observed at high resolution with \hrieuv, which is consistent with the `fireball' effect first detected by \citet{antolin2023extreme}. Signatures of this effect can also be seen in some of the AIA passbands.

Using the separation angle between AIA~171~\AA\, and \hrieuv at the time, we were able to calculate the total rain speed of downflows 1 and 2 to be $72.5 \pm 10.0$~km~s$^{-1}$ and $86.4 \pm 10.0$~km~s$^{-1}$, respectively. No clear acceleration or deceleration was observed, suggesting a balance of gas pressure and gravity. The speeds obtained are in agreement with previous studies \citep{DeGroof05, antolin2012observing, ishikawa2020temporal}. These dynamics are likely linked to the observed intensity brightening ahead of the rain suggesting compression. Compression from coronal rain (or prominence downflow) has been analysed analytically and numerically\citep{mackay2001evolution,AdroverGonzalez_2021AA...649A.142A}, predicting deceleration or piston-like oscillations around nodal points along the loop. \citet{oliver2014dynamics} have demonstrated that the abrupt formation of a condensation generates sounds waves upstream and downstream, that modify the pressure as they travel and effectively decelerates the clump and leads to a constant velocity during most of the downfall. Recently, this effect has been explained by \citet{hiller_2025A&A...696A.231H} as a dynamic mass contribution from the compressed material ahead of the rain, which then modifies the momentum of the rain. 

Based on the EUV absorption feature observed by \hrieuv, we calculated the width of downflow~1 to be $549\pm27$~km. This width falls in line with those found with the same instrument at similar resolution by \citep{antolin2023extreme}, and off-limb with SJI~2796~\AA, 1400~\AA, and 1330~\AA\, at similar resolution with IRIS by \citep{antolin2015multi}, strengthening the previous result that coronal rain widths are independent of temperature. As in \citet{csahin2023spatial}, the width of of the clump was observed to be constant along the loop's length. This property remains unexplained, and may be attributed to the fundamental scales over which coronal heating operates \citep{antolin2022multi} and/or to the pressure balance that follows thermal instability. 

The length of the EUV absorption feature created by downflow 1 is  $1.1 \pm 0.3$ Mm, which is similar to the width of the cool (chromospheric) coronal rain core found in the numerical simulations by \citep{Antolin_2022ApJ...926L..29A}. The actual length of the coronal rain clump (extending to higher transition region temperatures) is most likely greater as the tail can be elongated \citep{Fang_2013ApJ...771L..29F}. Also, we believe that downflow~1 is a collection of clumps, as suggested by the features seen in AIA~304~\AA. To refine our length estimate, we multiplied the speed of downflow 1 by the time interval between the impact and the onset of the upflow (matched by the continuous intensity enhancement observed during that time), to obtain a value of  $8.7 \pm 1.2$Mm. The studies referred to above report a broad distribution of rain lengths, from a few Mm up to 20 Mm. \citet{Fang_2015ApJ...807..142F} found that coronal rain lengths are subjected to shear flows, gas pressure and gravity as they descend.

{Several works have shown that gas pressure ahead of the rain is the leading cause of its deceleration, and mostly constant velocity at the time of impact. \citet{oliver2014dynamics, MartinezGomez_2020AA...634A..36M} have shown that the terminal speeds depend on the density contrast between the rain clump and the loop's density. Recently, \citet{Tan_2023MNRAS.520.2571T} in the context of cool clouds accreting towards galactic centres in the circumgalactic or intracluster medium, approximate the accretion braking process in the limit of small hydrodynamic drag as:}
\begin{equation}
    v=\frac{m}{\dot{m}}(g_{\text{eff}}-\dot{v}),
\end{equation}
{where we have modified their equation to account for the effective gravity $g_{\text{eff}}$ along a curved loop. Taking the average of $g_{\text{eff}}=0.171$~km~s$^{-2}$ for an elliptic loop with ratio of loop height to half baseline between 0.5 and 2, we have:}
\begin{equation}
    t_{\text{grow}}=\frac{v_{CR}}{g_{\text{eff}}},
\end{equation}
{where $v_{CR}=72.5~$km~s$^{-1}$ is the final velocity we find for downflow 1, and $t_{\text{grow}}$ is the growing time of the rain. We find, $t_{\text{grow}}\approx7~$min, which corresponds well with the time over which the EUV absorption of downflow 1 is observed (see e.g. Fig.~\ref{fig:131_td_splines})}, and roughly half the time where we estimate downflow 1 to occur.

In our study, the compression produced by the rain {is observed to be mostly isothermal while compressing the plasma to a maximum factor of 1.4. Based on the advective timescale being shorter than the conductive and radiative timescales, we assumed adiabatic conditions during the compression and} calculated the ratio of specific heats to obtain a value of $\gamma=1.085 \pm 0.218$. {It is possible, however, that the compression just beneath the rain is unresolved, with a much stronger radiative region than currently observed.} Also, {from the accretion braking process} we expect an important transfer of mass and momentum from the compressed region to the rain. {\citet{hiller_2025A&A...696A.231H} have shown that the numerical simulation results of rain downflows of \citet{oliver2014dynamics, MartinezGomez_2020AA...634A..36M} can be explained with the model of a virtual mass ahead of the rain}. The energy of the compression would then go to slow down the cooling from the rain and perhaps reionise it as it falls, thereby explaining the isothermal behaviour of the compressed plasma. \citet{hiller_2025A&A...696A.231H} {obtain the following relation for the falling speed:}
\begin{equation}
    v_{CR}=-\frac{g_{\text{eff}}H_{\text{eff}}/c_s}{1+\frac{2DH_{\text{eff}}}{F_bh_b}\sinh(\tau_{\text{eff}})}\tau_{\text{eff}},
\end{equation}
where we have also modified their equation to account for the effective gravity stated above, $H_{\text{eff}}$ is the effective pressure scale height for our loop at 1~MK, $F_b$ is the density contrast between the rain clump and the loop, $h_b$ is the length of the rain clump ahead of downflow 1, which we estimated to be $1.1\pm0.5$~Mm, where the uncertainty includes the possibility of the rain being composed of smaller clumps. $D=0.6$ as given by \citet{hiller_2025A&A...696A.231H}, and $\tau_{\text{eff}}=c_s t/H_{\text{eff}}$ is the time during which downflow 1 occurs ($=15\pm2$~min). Replacing with our values we obtain total speeds of $v_{CR}=79.4\pm17.7$~km~s$^{-1}$. In this calculation the main uncertainty is the length of the rain clump. Also, it is unclear how the assumed structuring (into smaller clumps) would affect on the dynamics. Despite the uncertainty, it is notable that the obtained rain speeds from accretion braking are very close to the observed total speed of the rain. \citet{Kohutova_2017AA...602A..23K} have conducted 2.5D-MHD simulations of coronal rain where, in addition to the gas pressure gradients, they show that magnetic pressure and tension can also play a role in decelerating the condensations. Density ratios (condensation to ambient) of 100 lead to piston-like effects where the condensations stop and bounce up and down prior to impact. High magnetic field strengths exacerbate this effect due to the bending of the field lines in the lower atmosphere. Only very high density ratios above 1000 (or even 10,000) effectively remove the piston effect. In our observations the density ratio is on the order of 170 and yet we do not observe this piston effect. The discrepancy may lie in the difference between 2.5D and 3D. Also, the boundary conditions implemented in the numerical models for the lower atmosphere can play a crucial role in the coronal rain dynamics.

{Furthermore, we note that the faster speeds of downflow~2 are also consistent with accretion braking, since once the first clump falls the upstream material is more rarefied, leading to less compression and less transfer of mass and energy to the second clump (which is often observed in coronal rain). These results suggest that coronal rain can be used as a template for accretion braking, which is not possible in observations of the circumgalactic or intracluster media.}

{While the adiabatic assumption may not hold, it is very likely that the $\gamma$ values are still significantly lower than the ideal gas case because of the presence of neutrals. Similarly} low $\gamma$ values under the adiabatic assumption, have been reported previously by \citet{van2011first}, who found $\ \gamma=1.10\pm0.02$. However, they observe a warm, highly ionised coronal loop using spectroscopic data from Hinode/EIS, and therefore interpret their results based on a very strong thermal conduction. Similarly,  \citet{prasad2018polytropic} also report low $\gamma$ values. Their research studied 30 active region loops using a DEM analysis and found a temperature dependence on $\gamma$, with cooler coronal loops exhibiting lower values. \citet{Vashalomidze_2019Ap.....62...69V} calculate $\gamma$ at the time of coronal rain onset, and found values of $2.1\pm0.11$, which they interpret as signatures of thermal instability. In our case, a low $\gamma$ value {is also linked to the structural complexity of the plasma, and may also} suggest the presence of molecules. The EUV absorption already indicates the presence of neutral hydrogen and helium and therefore temperatures likely under 10,000~K, {probably closer to 5,000~K}. Based on the spectral line width of H$\alpha$ with SST observations, \citet{antolin2015multi} find upper thresholds for the rain temperatures of 5000~K. \citet{Mulay_2021MNRAS.504.2842M} and \citet{Mulay_2023MNRAS.518.2287M} have found molecular H$_2$ emission in flare ribbons and jets, indicating temperatures around 4000~K. Hence, it is possible that molecules such as H$_2$ are also formed in coronal rain. 

We also detect upward motions that follow the impact of downflows 1 and 2. Rebound flows and shocks produced by the rain's impact have been reported first by \citet{antolin2023extreme} with \hrieuv. In our case, the upward motions are only clearly observed in the AIA~193~\AA\, and 211~\AA\, passbands, indicating temperatures of $\log T=6.1-6.3$, supported by the DEM analysis. The upward motions travel at a constant total velocity of $(85.1 - 86.5) \pm 10.0 $~km~s$^{-1}$. Since these speeds are much lower than the sound speed we interpret them as hot rebound flows, which then start to reheat the loop, indicating the start of a new TNE-TI cycle. Such atmospheric responses are expected from numerical simulations \citep[e.g.][]{muller2003dynamics, Antolin_2022ApJ...926L..29A}, and have been shown to affect the subsequent thermal evolution of the loop \citep{Johnston_2025ApJ...982..131J}. These upward motions can be traced clearly up to the formation height of downflow 1. The upflows are seen only about a minute after the downflow impact, further supporting the presence of multiple clumps accompanying downflows~1 and 2. The widths of upflow~1 is $\approx860~$km, {more than 50\%} larger than the rain width. This could be explained by the lower spatial resolution of AIA. However, an additional effect may contribute to these relatively large widths. It is plausible that the rain's impact generates transverse magnetic perturbations in the low transition region, due to the higher plasma $\beta$ at those heights, which, in turn, should generate compression in the neighboring field lines, leading to upward flows in a wider volume than that traced by the rain. The future higher resolution observations in hot lines offered by MUSE and EUVST should be able to confirm this hypothesis.

One of the most visible effects of the rain in the present case is its impact on the lower atmosphere. The rain shower, composed of small clumps, produces strong and rapid variability on the order of minutes observed in all AIA EUV channels, in \hrieuv and the SJI~1400~\AA\, channel. \citet{Reale_2013Sci...341..251R} report prominence fallback from very large heights to produce similarly strong intensity brightenings across all EUV wavelengths. However, they also report emission in the AIA~UV (low chromospheric) wavelengths, which are likely due to the higher falling speeds (up to 450~km~s$^{-1}$, probably due to the negigible effect of the magnetic field in their case) and consequently larger kinetic energies. The strong brightenings we observe are accompanied by a more steady and lower intensity enhancement over tens of minutes, matching roughly the duration of the rain shower, visible in AIA~131~\AA, 171~\AA\, and \hrieuv. This variation of intensity is likely due to the inhomogeneity present in the rain. The fact that the impact is seen in SJI 1400~\AA\, {for which the intensity increase is the strongest (compared to \hrieuv and AIA~193~\AA)}, indicates that the condensations reach the low transition region {and deposit most of their energy there}. The width and location of the brightening caused by the impact is very similar between SJI~1400~\AA, \hrieuv and AIA~193~\AA, which suggests that the transition region thickness is very thin (below $\approx1200$~km), matching predictions from numerical modelling \citep[e.g.][]{reale2014coronal}. {During the impact, all hot components become narrower, which is expected from the} heating.

We estimated the duration of downflow~1 to be  $\Delta_{CR}\approx3$~min, with a kinetic energy of {$(6.18 \pm 1.8)\times 10^{25}$}~erg. The energy of the rain shower $K_{SH}$ can be written as:
\begin{eqnarray}    
K_{SH} &=& K_{CR}N_{clumps} \\
 &\approx& K_{CR}\left(\frac{\Delta t_{SH}}{\Delta t_{CR}}\right) \label{eq:Ksh},
\end{eqnarray}
{where $N_{clumps}$ is the total number of rain clumps and $\Delta t_{SH}$ is the total duration of the rain shower, which we estimated to be roughly 35~min. In Eq.~\ref{eq:Ksh} we have made the reasonable assumption that the length of an individual clump times $N_{clumps}$ should be roughly equal to the length of the shower.} According to the time-distance diagrams (notably AIA~304~\AA), we can see that the bulk of the rain shower lasts 15~min (where the EUV absorption is predominant in AIA~304~\AA). Hence, the total energy of the rain shower is roughly {$4.64\times10^{26}$}~erg, that is, on the order of a microflare. {This is likely a lower bound since in the calculation of the rain's density we assumed a temperature of 10,000~K for the rain, which is likely too high}. Finally, the length of the rain is inferred rather than directly observed. In any case, these assumptions should compensate each other to some extent. 

The significant energy of the impact explains its visibility in all AIA/EUV wavelengths. We also found that {roughly} $15\%$ of the clump's kinetic energy is transferred to the thermal and kinetic energy of the rebound flow, along with a negligible wave energy. Hence, the majority of the energy is radiated away upon impact in the transition region. It is interesting to make the analogy to the `precipitation limit' feedback mechanism observed at the much larger scales of the intracluster medium (ICM). In that field, cool cores containing relatively dense and very cold gas are observed to accrete towards the supermassive black holes at the  centres of galaxy clusters \citep{Voit_2015Natur.519..203V}. These dense inflows impart momentum to the accretion disks, which power energetic AGN jets that subsequently heat the ICM and can slow down the cooling and cool core generation. Hence, a delicate balance is achieved between the number of cool core accretions and the temperature of the ICM from the AGN jets. These `precipitation limit' cycles with the cool cores are analogous to the TNE cycles with coronal rain found in the solar corona. In this work, this analogy is further extended since we show observationally that the rebound flow from coronal rain can have a major impact on the hosting coronal loop, not only through increasing the temperature but also modifying its morphology. 

Lastly, we observe a continuous intensity enhancement at the footpoint of the loop hosting the rain in SJI~1400~\AA, \hrieuv and AIA~193~\AA\, during the entire observation sequence, which could correspond to the previously reported steady character of the moss \citep{Antiochos_2003ApJ...590..547A, Brooks_2009ApJ...705.1522B}. These intensity enhancements are lower amplitude but have similar widths as those observed during impact,  ({especially the broad components with widths between 2~Mm and 10~Mm}). This suggests a steady footpoint heating component of similar heating scale length, which could be the kind predicted by TNE-TI. We can estimate or place a lower bound to the volumetric heating of this component with the following reasoning. The kinetic energy of coronal rain is very likely close to the total integrated volumetric heating from coronal heating over the duration prior to the rain. Although a proper estimation of this hypothesis awaits with numerical modelling, this can be justified by the fact that the footpoint heating in the low corona is partly lost in the transition region being radiated away, but a large part of it goes into producing a mild chromospheric evaporation that increases the coronal density \citep[leading to scaling laws such as RTVS,][]{Serio_1981ApJ...243..288S}. Most of this density collapses into the condensation, whose potential energy is then converted into kinetic energy as it falls. If so, we can write:
\begin{equation}
    E_{CH}\gtrapprox K_{SH},
\end{equation}
where $E_{CH}$ denotes the total integrated coronal heating from the beginning of the TNE-TI cycle in which the coronal rain forms, and {$K_{SH}$} is the kinetic energy of the rain shower, estimated to be {$4.64\times10^{26}$}~erg over its 15~min duration. On the other hand, the coronal heating energy is given by:
\begin{equation}
    E_{CH}=\int_0^{\tau_{TNE}} \int_0^{s_H} 2H_{CH}~dAdt,
\end{equation}
where $H_{CH}$ is the unknown, roughly steady volumetric heating rate; $s_H$ is the heating scale length, which we assume to be {set by the width of the observed persistent broad hot components at the loop footpoint, between 2~Mm and 10~Mm}; the factor of 2 is to account for both footpoints assuming similar heating conditions\footnote{Numerical simulations indicate that coronal rain often falls towards the footpoint with lower heating rate due to the lower pressure at the footpoint \citep{Froment_2018ApJ...855...52F}}; and $\tau_{TNE}$ is the period of the cycle. Unfortunately, this observational sequence is not long enough to observe another coronal rain event in the same loop with \hrieuv. However, looking further back in time with AIA~304~\AA\, reveals another coronal rain event in the same loop at 22:00UT on October 31st, that is, roughly 2 hrs prior to the studied event. We can take the area $A$ to be set by the width of the upward flow at the footpoint (assuming a circular cross-section), and since this area is very similar to that of the rain clumps, we have:
\begin{equation}
    H_{CH}\gtrapprox\frac{K_{SH}}{2s_{h
}A_{SH}\tau_{TNE}}=\frac{\rho_{CR}v_{CR}^3}{4s_H}\left(\frac{w_{CR}}{w_{SH}}\right)^2\frac{\Delta t_{SH}}{\tau_{TNE}},
\end{equation}
Replacing by the numerical values, we get a volumetric heating rate  {$H_{CH}\gtrapprox(0.41 - 2.05)\times10^{-2}~$erg~cm$^{-3}$~s$^{-1}$, or $H_{CH}\gtrapprox10^{-2.04\pm0.35}$~erg~cm$^{-3}$~s$^{-1}$}. Recently, \citet{ishigami2024spectroscopic} analysed spectroscopically several coronal loops with Hinode/EIS and found heating scale heights of $s_H=10\pm4~$Mm with a volumetric heating rate of $10^{-2\pm0.5}~$erg~cm$^{-3}$~s$^{-1}$. {Both, the heating scale height and} the volumetric heating rate that we find are very close to those reported. Although several other factors can influence TNE and coronal rain formation, this similarity also suggests that the coronal loops studied in \citet{ishigami2024spectroscopic} may be in a state of TNE. More generally, our study illustrates how coronal rain observations can be used to infer key properties of coronal heating. 

\section*{Acknowledgements}
{We would like to thank Ish Kaul, Peng Oh and Max Gronke for valuable discussions on accretion braking. Similarly, we would like to thank the anonymous referee for the many valuable comments that greatly improved the quality of this manuscript.} This research received support by the International Space Science Institute (ISSI) in Bern, through ISSI International Team project \#545 (`Observe Local Think Global: What Solar Observations can Teach us about Multiphase Plasmas across Physical Scales'). Data are courtesy of Solar Orbiter, SDO and IRIS. Solar Orbiter is a space mission of international collaboration between ESA and NASA, operated by ESA. The EUI instrument was built by CSL, IAS, MPS, MSSL/UCL, PMOD/WRC, ROB, LCF/IO with funding from the Belgian Federal Science Policy Office (BELPSO); Centre National d'Études Spatiales (CNES); the UK Space Agency (UKSA); the Deutsche Zentrum f\"ur Luftund Raumfahrt e.V. (DLR); and  the Swiss Space Office (SSO). The building of EUI was the work of more than 150 individuals during more than 10 years. We gratefully acknowledge all the efforts that have led to a successfully operating instrument. SDO is a mission for NASA's Living With a Star (LWS) program. IRIS is a NASA small explorer mission developed and operated by LMSAL with mission operations executed at NASA Ames Research Center and major contributions to downlink communications funded by ESA and the Norwegian Space Centre. All images in this manuscript have been made with Matplotlib on Python \citep{Hunter_2007}. This research used version 7.0 of the SunPy open source software package \citep{sunpy_community2020}. 

\section*{Data Availability}
The Solar Orbiter \hrieuv data used in this articles is part of Data Release 6.0 (\url{https://doi.org/10.24414/z818-4163}), which is publicly available. Similarly, the AIA data and IRIS data are publicly available and available at \url{https://iris.lmsal.com}.

%TC:endignore

\bibliographystyle{mnras}
\bibliography{mnras}
\bsp	% typesetting comment
\label{lastpage}
\end{document}